\documentclass[prb,twocolumn,amsmath,amssymb,superscriptaddress,longbibliography]{revtex4-2}

\usepackage{color,times}
\usepackage{dcolumn}
\usepackage{multirow}
\usepackage{amssymb,amsfonts,amsmath,graphicx}
\usepackage{epsfig}
\usepackage{xcolor}
\usepackage{SIunits}
\usepackage{braket}
\usepackage{epstopdf}
\usepackage[utf8]{inputenc}
\usepackage[english]{babel}
\definecolor{darkblue}{rgb}{0, 0, 0.8}
\usepackage[colorlinks=true, breaklinks=true, linkcolor=darkblue, citecolor=darkblue, urlcolor=darkblue]{hyperref}

\graphicspath{{..//figures//}}

\newcommand{\dd}{\mathrm{d}}

\begin{document}
	
	\title{Continuous Symmetry Breaking in a Two-dimensional Rydberg Array}
	\author{Cheng~Chen$^*$}
	\affiliation{Universit\'e Paris-Saclay, Institut d'Optique Graduate School,\\
		CNRS, Laboratoire Charles Fabry, 91127 Palaiseau Cedex, France}
	\author{Guillaume~Bornet$^*$}
	\affiliation{Universit\'e Paris-Saclay, Institut d'Optique Graduate School,\\
		CNRS, Laboratoire Charles Fabry, 91127 Palaiseau Cedex, France}
	\author{Marcus~Bintz$^*$}
	\affiliation{Department of Physics, University of California, Berkeley, California 94720 USA}
	\author{Gabriel~Emperauger$^*$}
	\affiliation{Universit\'e Paris-Saclay, Institut d'Optique Graduate School,\\
		CNRS, Laboratoire Charles Fabry, 91127 Palaiseau Cedex, France}
	\author{Lucas~Leclerc}
	\affiliation{Universit\'e Paris-Saclay, Institut d'Optique Graduate School,\\
		CNRS, Laboratoire Charles Fabry, 91127 Palaiseau Cedex, France}
	\affiliation{PASQAL SAS, 7 Rue Leonard de Vinci, 91300 Massy, France }
	\author{Vincent~S.~Liu}
	\affiliation{Department of Physics, University of California, Berkeley, California 94720 USA}
	\author{Pascal~Scholl}
	\affiliation{Universit\'e Paris-Saclay, Institut d'Optique Graduate School,\\
		CNRS, Laboratoire Charles Fabry, 91127 Palaiseau Cedex, France}
	\affiliation{California Institute of Technology, Pasadena, CA 91125, USA}
	\author{Daniel~Barredo}
	\affiliation{Universit\'e Paris-Saclay, Institut d'Optique Graduate School,\\
		CNRS, Laboratoire Charles Fabry, 91127 Palaiseau Cedex, France}
	\affiliation{Nanomaterials and Nanotechnology Research Center (CINN-CSIC), 
		Universidad de Oviedo (UO), Principado de Asturias, 33940 El Entrego, Spain}		
	\author{Johannes~Hauschild}
	\affiliation{Department of Physics, University of California, Berkeley, California 94720 USA}
	\affiliation{Department of Physics, Technical University of Munich, 85748 Garching, Germany}
	\affiliation{Munich Center for Quantum Science and Technology (MCQST), Schellingstr. 4, D-80799 M\"unchen, Germany}
	\author{Shubhayu~Chatterjee}
	\affiliation{Department of Physics, University of California, Berkeley, California 94720 USA}
	\author{Michael~Schuler}
	\affiliation{Institut f\"ur Theoretische Physik, Universit\"at Innsbruck, A-6020 Innsbruck, Austria}
	\author{Andreas~M.~L\"auchli}
	\affiliation{Laboratory for Theoretical and Computational Physics, Paul Scherrer Institute, 5232 Villigen, Switzerland}
	\affiliation{Institute of Physics, Ecole Polytechnique F\'ed\'erale de Lausanne (EPFL), 1015 Lausanne, Switzerland}
	\affiliation{Institut f\"ur Theoretische Physik, Universit\"at Innsbruck, A-6020 Innsbruck, Austria}
	\author{Michael~P.~Zaletel}
	\affiliation{Department of Physics, University of California, Berkeley, California 94720 USA}
	\affiliation{Materials Sciences Division, Lawrence Berkeley National Laboratory, Berkeley, CA 94720, USA}
	\author{Thierry~Lahaye}
	\affiliation{Universit\'e Paris-Saclay, Institut d'Optique Graduate School,\\
		CNRS, Laboratoire Charles Fabry, 91127 Palaiseau Cedex, France}
	\author{Norman~Y.~Yao}
	\affiliation{Department of Physics, University of California, Berkeley, California 94720 USA}
	\affiliation{Materials Sciences Division, Lawrence Berkeley National Laboratory, Berkeley, CA 94720, USA}
	\affiliation{Department of Physics, Harvard University, Cambridge, Massachusetts 02138 USA}
	\author{Antoine~Browaeys}
	\affiliation{Universit\'e Paris-Saclay, Institut d'Optique Graduate School,\\
		CNRS, Laboratoire Charles Fabry, 91127 Palaiseau Cedex, France}

	\date{\today}
	
	\maketitle
	
	\textbf{Spontaneous symmetry breaking underlies much of our classification of phases of matter 
		and their associated transitions~\cite{landauTheoryPhaseTransitions1937, 
			landauTheorySuperconductivity1950, keplerNiveSexangula1611}. 
		The nature of the underlying symmetry being broken determines many of the 
		qualitative properties of the phase; this is 
		illustrated  by the case of discrete versus continuous symmetry breaking.
		Indeed, in contrast to the discrete case, the breaking of a continuous symmetry 
		leads to the emergence of gapless 
		Goldstone modes controlling, for instance, the thermodynamic stability of the ordered 
		phase~\cite{goldstoneFieldTheoriesSuperconductor1961, 
			tasakiPhysicsMathematicsQuantum2020}.
		Here, we realize a two-dimensional dipolar XY model -- which exhibits a continuous 
		spin-rotational symmetry -- utilizing a programmable Rydberg quantum simulator. 
		We demonstrate the adiabatic preparation of correlated low-temperature states of 
		both the XY ferromagnet and the XY antiferromagnet. 
		In the ferromagnetic case, we characterize the presence of long-range XY order, a 
		feature prohibited in the absence of long-range dipolar interaction.
		Our exploration of the many-body physics of XY 
		interactions complements recent works utilizing the Rydberg-blockade mechanism to realize 
		Ising-type interactions exhibiting discrete spin rotation symmetry~\cite{Schauss2015,Bakr2018,Scholl2021, Ebadi2021}.}
	
	Constraints on when and how symmetries can be broken in many-particle systems abound.
	For example, long-wavelength fluctuations preclude the breaking of continuous symmetries 
	in low-dimensional systems with short-range interactions~\cite{blochZurTheorieFerromagnetismus1930, 
		peierlsQuelquesProprietesTypiques1935, merminAbsenceFerromagnetismAntiferromagnetism1966,  
		hohenbergExistenceLongRangeOrder1967a, brunoAbsenceSpontaneousMagnetic2001a}. 
	The presence of long-range interactions qualitatively alters this picture~\cite{Defenu2021}.
	On the one hand,
	they can stabilize certain forms of finite-temperature order, 
	which would otherwise be forbidden~\cite{dysonExistencePhasetransitionOnedimensional1969, 
		kunzFirstOrderPhase1976a, maleevDipoleForcesTwodimensional1976, frohlichPhaseTransitionsReflection1978}.
	On the other hand, they can also lead to frustration, where interactions compete with one another, 
	preventing the formation of order~\cite{diepFrustratedSpinSystems2013, castelnovoMagneticMonopolesSpin2008, 
		yaoQuantumDipolarSpin2018, 
		kelesAbsenceLongRangeOrder2018, kelesRenormalizationGroupAnalysis2018}. 
	Even when order persists in both the short- and long-range cases, the nature of this order, 
	including the dispersion of excitations or the decay of correlation functions, 
	can be fundamentally distinct~\cite{maleevDipoleForcesTwodimensional1976,DeBell2000,Taroni2008,Peter2012}. 
	
	Synthetic quantum systems are ideally suited to study these features. 
	While ultra-cold atoms in optical lattices have investigated continuous symmetry breaking 
	with contact interaction~\cite{Mazurenko2017}, dipolar molecules 
	in lattices~\cite{Yan2013, Christakis2022, Chomaz2022, Leo2018} or 
	trapped ions \cite{Richerme2014, Jurcevic2014,Maghrebi2017,Feng2022} are promising platforms to realize the long-range case.
	Here, we use a Rydberg quantum simulator to realize a long-range interacting, two-dimensional XY spin system 
	with either ferromagnetic (FM) or antiferromagnetic (AFM) couplings.
	We arrange up to $N=100$ dipolar interacting Rydberg atoms into a defect-free square lattice, 
	so that the many-body ground state in either the FM or AFM case is in a continuous symmetry breaking phase 
	characterized by off-diagonal long-range order~\cite{yangConceptOffDiagonalLongRange1962}.
	For the dipolar XY FM, theory predicts that this continuous symmetry breaking order persists in the presence of thermal 
	fluctuations~\cite{kunzFirstOrderPhase1976a, maleevDipoleForcesTwodimensional1976, Peter2012}.
	On the contrary, dipolar interactions are 
	insufficient to stabilize finite temperature, 
	long-range order in the antiferromagnet~\cite{brunoAbsenceSpontaneousMagnetic2001a}. 
	Rather, one expects power-law decaying, algebraic long-range order due to Berezinskii-Kosterlitz-Thouless 
	physics~\cite{berezinskiiDestructionLongrangeOrder1971, berezinskiiDestructionLongrangeOrder1972, 
		kosterlitzOrderingMetastabilityPhase1973, kosterlitzCriticalPropertiesTwodimensional1974, 
		giachettiBerezinskiiKosterlitzThoulessPhaseTransitions2021a}.
		
\begin{figure*}
		\includegraphics{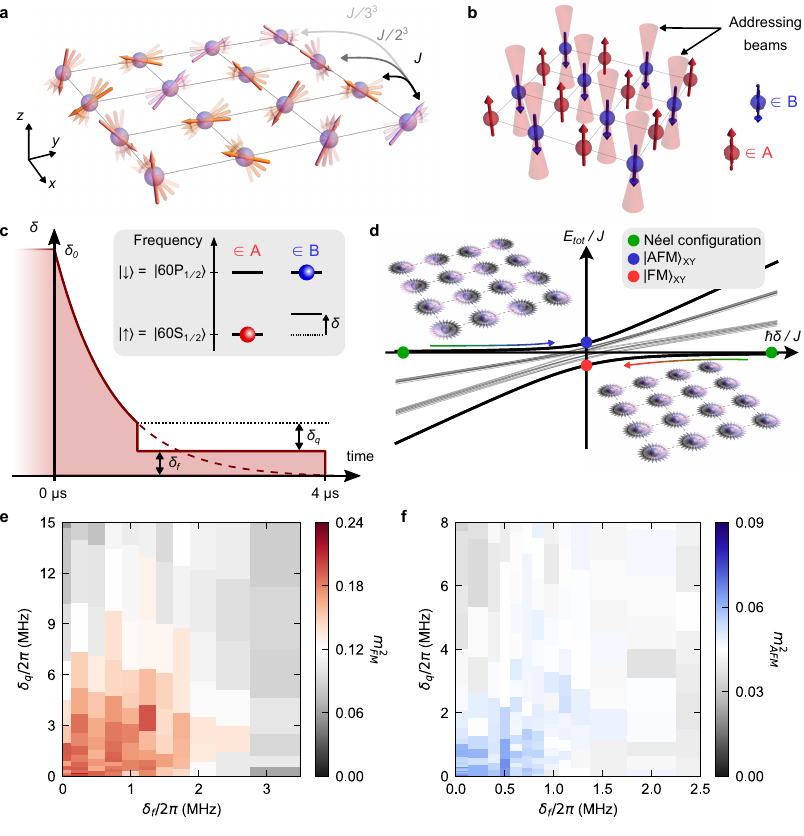}
		\caption{\textbf{Dipolar XY model in a Rydberg quantum simulator and experimental phase diagram.}
			\textbf{a}, Schematic depicting the long-range dipolar XY model. 
			An effective spin is encoded in a pair of Rydberg states which exhibit dipolar flip-flop interactions.
			\textbf{b}, A spatially dependent light-shift is used to prepare the system in a N\'eel spin configuration. 
			\textbf{c}, The amplitude $\delta$ of the light-shift is decreased as a function of time to a final value, 
			$\delta_f$. To study the robustness of the magnetic order with respect to an excess energy, 
			we introduce a diabatic quench of magnitude $\delta_q$.
			\textbf{d}, Energy spectrum of $H_{\textrm{tot}}$ as a function of $\delta$, for $N=2\times 3$. When starting in the ground state 
			for $\hbar \delta/J \gg 1$, the system is adiabatically ramped to the ferromagnetic XY state, 
			pictured by the colored fluctuating arrows correlated in directions. 
			When starting in the highest excited state for $\hbar \delta/J \ll -1$, the system 
			is adiabatically ramped to the antiferromagnetic XY state, portrayed by the anticorrelated fluctuating arrows. 		
			\textbf{e}, Ferromagnetic phase diagram depicting the magnetization squared as a function of the final 
			staggered field strength, $\delta_f$ and the diabatic quench magnitude, $\delta_q$. 
			Symmetry breaking is expected in a lobe about ($\delta_f=0, \delta_q=0$) and is destroyed by either 
			quantum ($\delta_f$) or thermal ($\delta_q$) fluctuations. On a 6 $\times$ 7 system, a crossover between 
			ordered and disordered behavior is observed. 
			\textbf{f}, Analogous phase diagram for the antiferromagnet. Note that  at finite temperature, 
			only algebraic long-range order is expected.}
		\label{fig:fig1}
	\end{figure*}

	Our main results are threefold.
	First, leveraging single-site addressing, we adiabatically prepare correlated low-temperature states of 
	both the XY FM and the XY AFM starting from a classical staggered spin configuration.  
	Second, we characterize the prepared states by measuring the full spatial profile of correlation functions. 
	In the ferromagnet, the system exhibits correlations consistent with the presence 
	of long-range order -- a feature prohibited in conventional short-range-interacting, two-dimensional 
	magnets~\cite{
		merminAbsenceFerromagnetismAntiferromagnetism1966, hohenbergExistenceLongRangeOrder1967a}.
	Meanwhile, in the antiferromagnet, correlations vanish at  long distances, 
	consistent with the decay expected from algebraic long-range order.
	We also show that the states produced are not classical FM or AFM. 
	Third, by introducing a partial quench into the adiabatic ramp, 
	we study the robustness of the magnetic order with respect to an excess energy akin to an effective temperature. 
	This allows us to probe the phase diagram of the dipolar XY model (Fig.~\ref{fig:fig1}).  
	
	%%%%%%%%%%%%%%%%%% Fig 1 setup %%%%%%%%%%%%%%%%%%%%
	%%%%%%%%%%%%%%%%%%%%%%%%%%%%%%%%%%%%%%%%%%%%

	The experimental setup (Fig.~\ref{fig:fig1}a) consists of a two-dimensional square lattice of $^{87}$Rb 
	atoms trapped in an optical tweezer array~\cite{Scholl2021}.
	We encode an effective spin $1/2$ in a pair of opposite-parity Rydberg states, 
	$\left|\uparrow\right\rangle = |60S_{1/2}\rangle$ and $\left|\downarrow\right\rangle = |60P_{1/2}\rangle$.
	Resonant dipole-dipole interactions between the spins naturally realize the %so-called 
	dipolar XY model~\cite{Browaeys2020}, 
	\begin{equation}\label{Eq:HXY}
	H_{\rm XY}= - {J\over 2}
	\sum_{i < j}  \frac{a^3}{r_{ij}^3}  (\sigma^x_i \sigma^x_j + \sigma^y_i \sigma^y_j),
	\end{equation}
	where $\sigma_i^\alpha$ are Pauli matrices, $r_{ij}$ is the distance between spins $i$ and  $j$,  
	$J/h= 0.77~$MHz is the dipolar interaction strength, and $a=12.5~\mu$m is the lattice spacing; 
	here, the quantization axis is defined by an external magnetic field perpendicular to the lattice plane, 
	which ensures that the dipolar interactions are isotropic.
	The Hamiltonian exhibits a continuous $U(1)$ symmetry corresponding to the conservation 
	of total $z$-magnetization, $M^z = \sum_i \sigma^z_i$ (see Methods Sec.~\ref{SubSM:Symmetry}).
	
	%%%%%%%%%%%%%%%% Figure 2 %%%%%%%%%%%%%%%%%%%%	
	\begin{figure*}[t!]
		\includegraphics{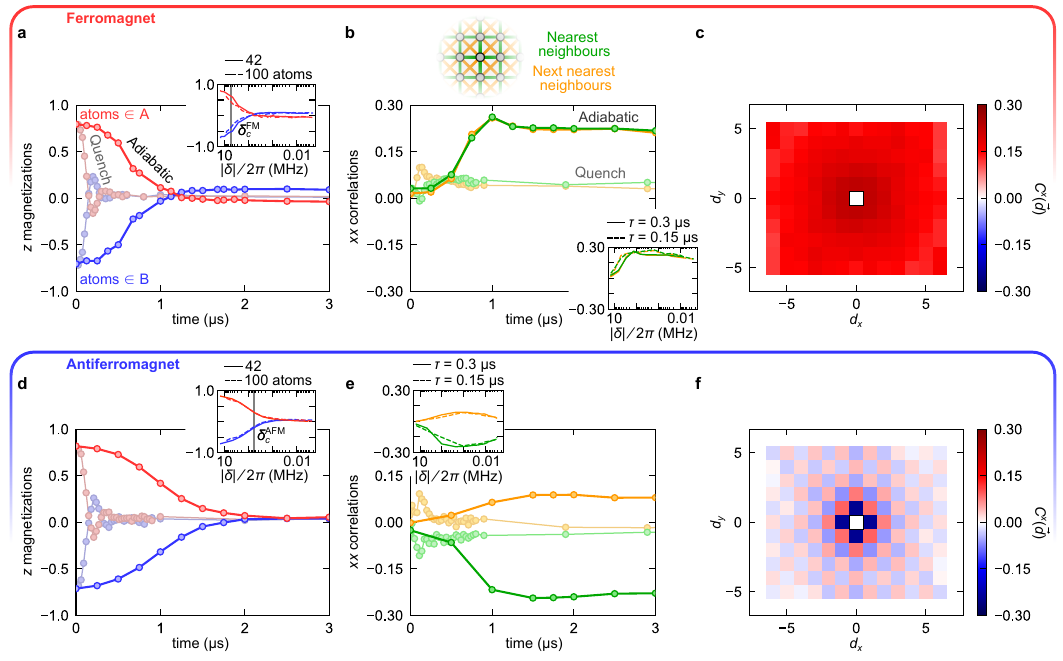}
		\caption{\textbf{Adiabatic preparation of dipolar XY ferro- and antiferromagnets.}
			\textbf{a}, Sublattice-resolved  magnetization $\langle \sigma^z_i \rangle$ as the staggered field $\delta(t)$ is  
			reduced.  At $t=0$, the state is prepared in a classical N\'eel state along the $z$-axis, as indicated by the 
			opposing magnetization of atoms in the A (red) and B (blue) sublattices.   As the staggered field $\delta(t)$  
			is turned off, either adiabatically or via a sudden quench, the N\'eel magnetization decays towards zero. 
			(inset) Comparison of the $z$-magnetizations decay as a function of $\delta$ for a $6\times 7$ 
			versus a $10\times 10$ lattice. The gray vertical line indicates the value $\delta_{\rm c}^{\rm FM}$ where the phase transition 
			occurs, inferred from the theory (Methods~\ref{SubSM:DMRG}).
			\textbf{b}, The formation of a low-energy XY-ferromagnet is detected via the in-plane 
			two-point correlation function, 
			$C^x_{i,j}$. Data is shown for $i, j$  averaged over either nearest or next-nearest pairs. 
			The sudden quench produces additional energy which destroys the XY order and leads to 
			correlations near zero. (inset) Nearest and next-nearest correlations for two different adiabatic ramp rates. 
			\textbf{c}, $xx$ correlations as a function of displacement, 
			$C^{x}(\vec{d}\hspace{0.5mm}) \equiv \langle  C^x_{\vec{r},\vec{r}+\vec{d}} \hspace{0.5mm}  \rangle_{\vec{r}}$, 
			measured at time, $t = 2~\mu$s (with $d_x$ and $d_y$ in units of lattice spacing $a$).
			\textbf{d-f}, Analogous results for the antiferromagnetic case.
			Crucially (\textbf{e,f}), we observe staggered correlations. }
		\label{fig:fig2}
	\end{figure*}
	%%%%%%%%%%%%%%%%%%%%%%%%%%%%%%%%%%%%%%%
	
	The starting point of our experiments is a classical  N\'eel spin
	configuration, i.e. a staggered arrangement  of spins $\ket{\downarrow}$ and $\ket{\uparrow}$ with 
	$M^z=0$, prepared in the following way
	(see Methods Sec.~\ref{SubSM:exp_seq}):  
	after initializing all the atoms in $\ket{\uparrow}$, we apply focused laser beams to produce spatially dependent light-shifts, 
	implementing  the Hamiltonian $H_{\rm Z} =  \hbar \delta \sum_i n_i$.  
	The $n_i$ form a staggered pattern with $n_i= 0$ on the A-sublattice and $n_i=(1+\sigma^z_i)/2$ 
	on the B-sublattice (Fig.~\ref{fig:fig1}b). We then sweep a global microwave pulse across the 
	resonance of the atoms in the A-sublattice  that flips their spin to $\ket{\downarrow}$. 
	This leads to the N\'eel configuration, which  is a good approximation of the ground state (for $\delta>0$) 
	or highest excited state (for $\delta<0$) of the total Hamiltonian 
	$H_{\textrm{tot}} = H_{\rm XY} + H_{\rm Z}$ for $\hbar|\delta| \gg J$.
	
	Starting from this
	configuration, we dynamically prepare highly-correlated, quantum many-body states by
	ramping down as a function of time the laser field producing the staggered light-shifts, either abruptly or 
	adiabatically (Fig.~\ref{fig:fig1}c) (for a discussion of an alternative preparation approach, 
	see Methods Sec.~\ref{SubSM:adiabatic_alternative}).
	In the adiabatic case, for $\delta(t) >  0$,  the ramp connects the N\'eel configuration to the  
	low-temperature ferromagnetic states of $H_{\rm XY}$, as shown in Fig.~\ref{fig:fig1}d. 
	Meanwhile, for $\delta(t) <  0$, 
	the adiabatic ramp prepares {\it negative} 
	temperature states of $H_{\rm XY}$ or equivalently, low-temperature antiferromagnetic states of 
	$-H_{\rm XY}$ (Fig.~\ref{fig:fig1}d)~\cite{sorensenAdiabaticPreparationManybody2010}.
	In the thermodynamic limit of both cases, a quantum phase transition is expected to 
	occur at some critical $\delta_c^{\textrm{FM/AFM}}$, between the N\'eel configuration and the XY order
	(Methods~\ref{SubSM:QPT}).

	To investigate the XY ferromagnet, we begin with a $6 \times 7$ lattice and utilize an exponential
	ramp profile, $\delta(t)\approx \delta_0 e^{- t/\tau}$, with $\delta_0 = 2\pi \times 15~$MHz and $\tau =0.3~\mu$s.
	As depicted in Fig.~\ref{fig:fig2}a, for both sublattices, the on-site $z$-magnetization, $2\sum_{i\in A/B} \langle \sigma^z_i\rangle/N$,
	obtained by averaging over many realizations of the experiment, decreases toward zero, 
	with a residual late-time offset arising from experimental imperfections (see Methods Sec.~\ref{SubSM:Decoherence}).
	This is consistent with the XY ferromagnet, which orders in the equatorial plane, but by itself, is insufficient to diagnose the phase. 
	Indeed, quenching the staggered light-shifts (in less than $100~$ns) leads to a near infinite temperature state, 
	which also exhibits a magnetization that rapidly relaxes to zero  (lighter curves, Fig.~\ref{fig:fig2}a). 
	
	The key characteristic of the XY ferromagnet 
	is only revealed upon measuring the correlation function, 
	$C^x_{ij} = \langle \sigma^x_i \sigma^x_j \rangle - \langle \sigma^x_i \rangle  \langle \sigma^x_j \rangle$
	(Methods~\ref{SubSM:Symmetry}).
	For the quenched state, the correlation functions remain near zero for all times, 
	consistent with high-temperature behavior (lighter curves, Fig.~\ref{fig:fig2}b). 
	The dynamics of the adiabatic protocol are markedly distinct -- both nearest-neighbor  and next-nearest-neighbor 
	correlations grow to a stable non-zero value at late times, indicative of order~\cite{yangConceptOffDiagonalLongRange1962}.
	By switching the sign of $\delta_0$, we also investigate the XY antiferromagnet.
	Both the $z$-magnetization (Fig.~\ref{fig:fig2}d) and the correlation functions (Fig.~\ref{fig:fig2}e) 
	exhibit qualitatively similar dynamics as the ferromagnetic case.
	One notable difference is that $C^x<0$ for nearest-neighbor correlations, indicating that neighboring spins have anti-aligned. 
	
	A few remarks are in order. First,  to explore the adiabaticity of our protocol, 
	we vary the time-constant of the exponential ramp.
	As shown in the insets of Fig.~\ref{fig:fig2}b,e, the dynamics of the correlation function  agree between 
	$\tau = 0.15~\mu$s and $\tau = 0.3~\mu$s, indicating that diabatic errors are not a limiting factor. 
	We confirm this by numerical simulation of the many-body dynamics (see Methods Fig.~\ref{fig:MPOMPS_energy}).
	Second, while the long-range tail of the dipolar interaction reinforces the XY FM order, it is weakly frustrating for the AFM~\cite{Defenu2021}.
	As a consequence, the phase transition between the N\'eel configuration and the 
	XY AFM is expected to occur at a smaller value of the staggered light-shift as compared 
	to the XY FM, i.e.~$|\delta_{\rm c}^{\textrm{AFM}}| < |\delta_{\rm c}^{\textrm{FM}}|$
	(see also Methods Sec.~\ref{SM:Adiabatic}). 
	This is indeed borne out by the data where we observe that the magnetization decays to zero faster as a function of $\delta$
	for the FM case than for the AFM.
	Third, we increase the system size to a $10 \times 10$ lattice and perform the analogous adiabatic preparation protocols.
	We find  the same behavior for all observables (insets, Fig.~\ref{fig:fig2}a,d), indicating that our results are robust 
	to finite-size effects~\cite{sandvikGroundstateParametersFinitesize1999}. 
	Finally, we observe that at the latest times, the correlations in both the FM and AFM cases exhibit a slow
	decay; we conjecture that this decay arises from a combination of residual atomic 
	motion and the finite lifetime of the Rydberg states (more details in Methods~\ref{SubSM:Decoherence}).  
	
	%%%%%%%%%%%%%%%%%%%%%%% Fig 3 %%%%%%%%%%%%%%%%%%
	\begin{figure*}[t!]
		\includegraphics{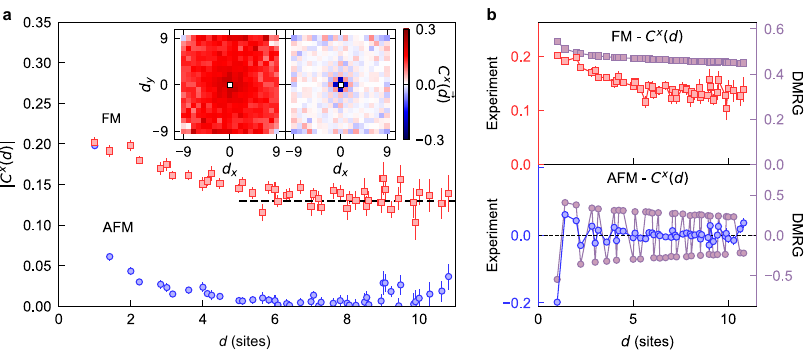}
		\caption{\textbf{Observing long-range XY order in a $10 \times 10$ lattice.} 
			\textbf{a},  xx correlations averaging over displacements of the same distance, $C^{x}(d)$. 
			The XY ferromagnet exhibits a plateau consistent with long-range order, while the  XY 
			antiferromagnet exhibits a decay to zero. 
			(inset) Spatial correlations as a function of displacement, measured at time $t = 1~\mu$s. 
			\textbf{b}, Comparison of the experimental data shown in \textbf{a} with the 
			ground-state results obtained from DMRG.}
		\label{fig:fig3}
	\end{figure*}
	%%%%%%%%%%%%%%%%%%%%%%%%%%%%%%%%%%%%%%%%%%%
	
	Our measurements of the local correlations suggest we have dynamically prepared low-temperature 
	states of the XY FM and AFM -- but are these states truly long-range ordered? 
	To investigate this, we measure the long-distance spin-spin correlations of the system after adiabatic preparation.
	In Fig.~\ref{fig:fig2}c,f [$6\times7$] and Fig.\,\ref{fig:fig3}a [$10\times10$]  
	we show the correlations  as a function of the displacement $\vec{d}$, averaging over initial positions:
	$C^{x}(\vec{d}\hspace{0.5mm}) \equiv \langle  C^x_{\vec{r},\vec{r}+\vec{d}} \hspace{0.5mm}  \rangle_{\vec{r}}$.
	The FM correlations are of constant sign  and appear to plateau at long distances, 
	indicative of long-range order, while the AFM correlations are staggered and exhibit a decay.
	For a more quantitative assessment, we focus on the $10 \times 10$ array and plot  $C^{x}(d)$, 
	averaging over displacements of the same distance $d = |\vec{d}\hspace{0.5mm}|$.  
	In the XY AFM, correlations decay to zero at large distances, indicating the absence of long-range order. 
	By contrast, the XY FM indeed exhibits a plateau, $C^{x}_\infty \sim 0.13$, which establishes 
	it as a magnetically ordered state with an effective magnetization 
	density $m_{\textrm{eff}} \equiv \sqrt{2 C^{x}_\infty} = 0.51$ (Methods~\ref{SubSM:Symmetry}).
	
	For additional insight, in Fig.~\ref{fig:fig3}b we compare the measured $C^{x}(d)$ against the exact 
	ground-state prediction obtained from density matrix renormalization group (DMRG) calculations 
	(see Methods Sec.~\ref{SubSM:DMRG})~\cite{whiteDensityMatrixFormulation1992, hauschildEfficientNumericalSimulations2018}.
	In the DMRG ground state, $C^{x}(d)$ does plateau in the FM, but slowly decays 
	in the AFM due to finite-size effects -- in the thermodynamic limit, both the FM and AFM ground states 
	are expected to be long-range ordered at zero temperature.
	While the qualitative structure of the measured $C^{x}(d)$ (e.g.~sign structure in the AFM case) 
	is consistent with theory, the experimental correlations are weaker. 
	A number of effects could contribute to this.
	For example, the finite fidelity of the initial  N\'eel  state introduces an entropy density 
	(i.e. an effective finite temperature). This is especially destructive to the AFM, 
	for which finite temperature long-range order is forbidden~\cite{brunoAbsenceSpontaneousMagnetic2001a,Defenu2021}, 
	in agreement with our observation.
	Other experimental imperfections including readout errors are discussed in the Methods, 
	Sec.~\ref{SM:exp_imperfections}; including these errors in our numerical simulations leads to 
	excellent agreement with the data for the $6 \times 7$ lattice (see Methods Fig.\,\ref{fig:mps_exp_comparison}).
	However, we also observe that running the adiabatic preparation protocol to longer timescales leads to additional 
	decoherence which adversely affects the ferromagnetic magnetization plateau in a non-trivial fashion; 
	in particular, correlations at the largest distances begin to 
	decay before their shorter-distance counterparts~(see Methods~\ref{SubSM:Decoherence}). 
	
	%%%%%%%%%%%%%%%%%%%%%%%%%%%%%%%%%%%%%
	\begin{figure}[t!]
		\includegraphics{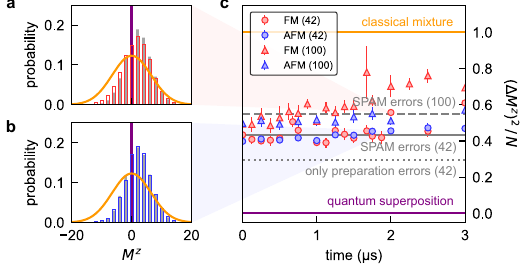}
		\caption{\textbf{Analysis of the $z$-magnetization  during the adiabatic ramp.} 
			Experimental histograms of the $z$-magnetization $M^z$ ($N =42$) for \textbf{a}, the FM and \textbf{b}, AFM case
			together with the ideal case (purple), and the expected distribution including state preparation and measurement errors (grey bars). 
			The orange line is the binomial distribution corresponding to a classical magnet (see text). 
			\textbf{c},  normalized variance $(\Delta M^z)^2/N$ as a function of time during the ramp, for the experiment 
			(circles for $N=42$, triangles for $N=100$), 
			the classical magnet (orange line) and 
			a perfect XY-magnet (purple line). Grey continuous and dashed lines: 
			ideal case including state preparation and measurement errors. 
			Dotted line: ideal case ($N=42$) including only state preparation errors. }
		\label{fig:fig4}
	\end{figure}
	%%%%%%%%%%%%%%%%%%%%%%%%%%%%%%%%%%%%
	
	As a final characterization of the prepared states, we investigate whether each 
	realization of the experiment produces a {\it classical} magnet pointing 
	in a random direction $\theta$ in the $xy$-plane
	or a genuinely {\it quantum} many-body state (see Methods \ref{SubSM:Symmetry}).
	To do so, we analyze the statistical distribution of $M^z$, which is conserved during the adiabatic ramp. 
	For a classical FM or AFM, each spin, aligned or anti-aligned along $\theta$, is 
	an equal superposition of $\ket \uparrow$ and $\ket \downarrow$, so that $M^z$ follows a binomial distribution. 
	By contrast, the ground state of $H_{\mathrm{XY}}$ is an eigenstate of $M^z$, and its variance should be zero.  
	Figure~\ref{fig:fig4}a,b shows experimental histograms of the $z$-magnetization
	at $t=2$\,$\mu$s for the FM and AFM. Figure~\ref{fig:fig4}c presents the variance for various times $t$. 
	We find that the states have a variance smaller than that of a binomial distribution, indicating that we do not prepare 
	classical magnets. 
	In fact, the measured non-zero variances can be fully 
	explained by the state preparation and measurement errors applied to the ideal distribution (see Methods \ref{SubSM:SPAM_errors}).  
	We have also checked the rotation invariance of the state around $z$ 
	by measuring the magnetization along $y$ and finding the same as along $x$. 
	Altogether, our measurements suggest a state which is a coherent quantum 
	superposition over a continuous family of classical configurations (see Methods~\ref{SubSM:Symmetry}). 
	For such a state, the defining signature of continuous symmetry breaking order is a long-distance plateau 
	in the correlation function $C^x(d)$ -- as we observed in the XY FM~\cite{tasakiPhysicsMathematicsQuantum2020}. 
	
	As mentioned earlier, the long-range order observed in the FM case should persist at finite temperature. 
	We therefore investigate the stability of the prepared magnetic orders as a function of an effective temperature.
	To do so, we insert a partial quench of amplitude $\delta_q$ into the ramp, 
	followed by an equilibration time of at least $1~\mu$s at a final value $ \delta_f$ of the staggered field (Fig.~\ref{fig:fig1}d): 
	the variable quench introduces excess energy into the system, and we observe a relaxation of the magnetization and correlations 
	during the equilibration time. %leading to the preparation of a higher-temperature state.
	We will use the amplitude of the quench, $\delta_q$, as a proxy for the final effective temperature (see Methods Sec.~\ref{SM:Therm}).
	After each $\{ \delta_f, \delta_q \}$ ramp, we measure the  in-plane magnetization squared,  
	$m^2_{\textrm{FM/AFM}} = \sum_{ij} (\pm 1)^{i+j} C^x_{ij}/N^2$ and construct the phase diagram shown in Figs.~\ref{fig:fig1}e,f.	
	Starting with the ferromagnet, for small values of  $\delta_f$ and  $\delta_q$ (corresponding to low effective temperatures), the 
	magnetization per site is of $\mathcal{O}(1)$, consistent with the ordered phase (Fig.~\ref{fig:fig1}e).
	As either $\delta_f$ or $\delta_q$ increases, the magnetization density decreases toward zero 
	indicating melting into a disordered phase. This is consistent with theoretical expectations, 
	where $\delta_q$  drives the transition via thermal fluctuations~\cite{kunzFirstOrderPhase1976a}, while $\delta_f$ 
	tunes across the quantum phase transition.
	We perform the same analysis for the antiferromagnet (Fig.~\ref{fig:fig1}f).
	Compared to the XY ferromagnet, we find that a much smaller region of the $\{ \delta_f, \delta_q \}$ 
	phase space exhibits significant AFM correlations, consistent with the frustration induced by the long-range interactions
	which destabilizes the phase. 
	
	{\it Outlook} -- Looking forward, our work opens the door to a number of future directions. 
	First, it would be interesting to investigate the nature of the phase transition between the 
	disordered and XY-ordered phases; this will require overcoming a number of technical 
	challenges including scaling to larger system sizes. Second, the ability to directly prepare 
	low-temperature states in different $M^z$ magnetization sectors suggests the possibility 
	of directly observing the so-called Anderson tower of states, which underlies continuous 
	symmetry breaking in finite quantum systems~\cite{andersonApproximateQuantumTheory1952, andersonBasicNotionsCondensed2010, 
		tasakiLongRangeOrderTower2019, beekmanIntroductionSpontaneousSymmetry2019a}; 
	the structure of these states has led to recent predictions for scalable spin squeezing by 
	quenching in the ferromagnetic XY phase~\cite{comparinRobustSpinSqueezing2022}.
	Finally, combining optical tweezer geometries which exhibit frustration (i.e.~triangular or kagome lattices) 
	with antiferromagnetic interactions leads to a rich landscape for exploring 
	frustrated magnetism and spin liquid physics~\cite{diepFrustratedSpinSystems2013, yaoQuantumDipolarSpin2018}. 
	
	\begin{acknowledgments}
		
		We acknowledge the insights of and discussions with M. Aidelsburger, L. Henriet, V. Lienhard, J. Moore, 
		C. Laumann, B. Halperin, E. Altman, B. Ye, E. Davis, and M. Block. 
		We are especially indebted to Hans Peter B\"uchler for insightful comments and 
		discussions about the role of dipolar-interactions in the XY model. 
		The computational results presented were performed in part using the FASRC Cannon cluster 
		supported by the FAS Division of Science Research Computing Group at Harvard University, 
		the Savio computational cluster resource provided by the Berkeley Research Computing program 
		at the University of California, Berkeley and the Vienna Scientific Cluster (VSC). 
		This work is supported by the European Union's Horizon 2020 research and innovation 
		program under grant  agreement No. 817482 (PASQuanS), 
		the Agence Nationale de la Recherche (ANR, project RYBOTIN and ANR-22-PETQ-0004 France 2030, project QuBitAF), and the 
		European Research Council (Advanced grant No. 101018511-ATARAXIA). 
		J.H.~acknowledges support from the NSF OIA Convergence Accelerator Program 
		under award number 2040549, and the Munich Quantum Valley, which is supported by 
		the Bavarian state government with funds from the Hightech Agenda Bayern Plus. 
		MS and AML acknowledge support by the Austrian Science Fund (FWF) through Grant No. I 4548.
		DB acknowledges support from MCIN/AEI/10.13039/501100011033 
		(RYC2018- 025348-I, PID2020-119667GA-I00, and European Union NextGenerationEU PRTR-C17.I1)
		M.Z. acknowledges support from the DOE Early Career program and  the Alfred P. Sloan foundation.  
		N.Y.Y.  acknowledges support from the Army Research Office (W911NF-21-1-0262), 
		the AFOSR MURI program (W911NF-20-1-0136),  the David and Lucile Packard foundation, 
		and  the Alfred P. Sloan foundation. 
		M.B. and V.L. acknowledge support  from NSF QLCI program (grant no.~OMA-2016245).
		S.C. acknowledges support  from the ARO through the MURI program (grant number W911NF-17-1-0323) 
		and from the U.S. DOE, Office of Science, Office of Advanced Scientific Computing Research, 
		under the Accelerated Research in Quantum Computing (ARQC) program.
	\end{acknowledgments}
	
	\section*{Author contributions}
	\noindent
	$^*$CC, GB, MB, and GE contributed equally to this work.
	CC, GB, GE, PS and DB carried out the experiments. 
	MB, LL, VSL, JH, SC and MS conducted the theoretical analysis and simulations. 
	AML, MPZ, TL, NYY and AB supervised the work.
	All authors contributed to the data analysis, progression of the project, 
	and on both the experimental and theoretical side. 
	All authors contributed to the writing of the manuscript. 
	Correspondence and requests for materials should be addressed to AB.
	
	\section*{Ethics Declaration}
	AB and TL are co-founders and shareholders of PASQAL.
	
%	\clearpage
	
%\bibliography{dipolarXY_combined_biblio}

	\clearpage
	
	\section*{Methods} 
	
	\subsection{Experimental methods}\label{SM:Exp_details}
	
	The realization of the dipolar XY model relies on our $^{87}\text{Rb}$ 
	Rydberg-atom tweezer array setup, 
	described in previous works~\cite{Barredo2016,Scholl2021}. 
	The pseudo-spin states are $\ket{\uparrow} = \ket{60S_{1/2}, m_J = 1/2}$ and 
	$\ket{\downarrow} = \ket{60P_{1/2}, m_J = -1/2}$. 
	We manipulate them using resonant microwaves at 16.7 GHz. 
	A $\sim50$-G magnetic field, perpendicular to the array, defines the quantization axis (Fig. \ref{fig:ExpSequ_SI}a)
	and shifts away the irrelevant Zeeman states of the $60S_{1/2}$ and $60P_{1/2}$ manifolds. 
	
	\subsubsection{Addressability in the tweezer array}\label{SubSM:Addressing}
	
	The addressing laser pattern used to prepare the initial classical N\'eel configuration  is generated by a 1013-nm laser beam 
	detuned from the transition between the intermediate state $6P_{3/2}$ and $\ket{\uparrow}$ (Fig.~\ref{fig:ExpSequ_SI}b). 
	The sign of the detuning sets the one of the light-shift: in the FM (resp. AFM) case, the frequency of the addressing laser 
	is tuned below (resp. above) the resonance by $\sim 250 ~$MHz.
	
	We use a dedicated spatial light modulator to produce the pattern of addressing beams.
	Each beam is focused on a $1/e^2$ radius of about $1.5~\mu$m, for a typical power of $60~$mW. 
	We measure the light-shift for each addressed atom by microwave spectroscopy 
	on the $\ket{\uparrow}-\ket{\downarrow}$ transition. 
	The average light-shift is $|\delta_0| = 2\pi \times 15~$MHz over the 42-atom array (21 addressed atoms), 
	and $|\delta_0| = 2\pi \times 9~$MHz 
	over the 100-atom array (50 addressed atoms). These values are dictated by available laser power. 
	For both arrays, the rms dispersion of $\delta_0$ across the addressing beams is $2.4\%$.
	
	\subsubsection{Experimental sequence}\label{SubSM:exp_seq}
	
	%%%%%%%%%%%%%%%%%%% Figure SM1 %%%%%%%%%%%%%%%%%%
	\begin{figure*}[t!]
		\includegraphics[width = 18.00 cm]{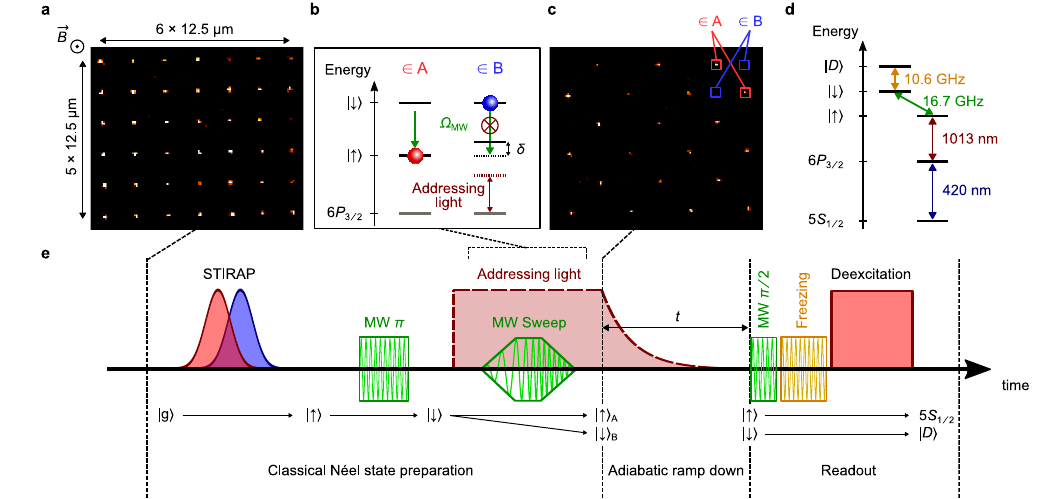}
		\caption{\textbf{Experimental procedures and sequence.}
			\textbf{a}, Fluorescence image of the atoms in a fully assembled $6 \times 7$ array.
			\textbf{b}, Scheme for the preparation of the initial staggered state.
			\textbf{c}, Detected staggered state, corresponding to the situation for which all the atoms in 
			sublattice A are in $\ket{\uparrow}$, 	and all the atoms in sublattice B are in $\ket{\downarrow}$.
			\textbf{d}, Schematics of the atomic level diagram.
			\textbf{e}, Experimental sequence.}
		\label{fig:ExpSequ_SI}
	\end{figure*}
	%%%%%%%%%%%%%%%%%%%%%%%%%%%%%%%%%%%%%%%%%%%
	
	The experimental sequence is shown in Fig.~\ref{fig:ExpSequ_SI}. After assembling the 
	array~\cite{Barredo2016} we use Raman sideband cooling along the radial directions of the tweezers, 
	and reach a temperature of $10\,\mu$K. We then optically pump the atoms in 
	$\ket{g}=\ket{5S_{1/2}, F = 2, m_F = 2}$ before adiabatically ramping down the tweezer
	depth by a factor $\sim 40$. Following this, we switch off the tweezers, and excite the atoms 
	to $\ket{\uparrow}$ using a two-photon stimulated Raman adiabatic passage (STIRAP) 
	with 421-nm and 1013-nm lasers ($\sim 2~\mu$s duration).
	
	To generate the  classical N\'eel configuration along $z$, we first transfer all the atoms from 
	$\ket{\uparrow}$ to $\ket{\downarrow}$ using a $54$~ns microwave $\pi$-pulse. 
	Subsequently, the addressing beams are applied to the atoms in sublattice B. 	
	We then transfer the atoms A from $\ket{\downarrow}$ back to $\ket{\uparrow}$ by an adiabatic 
	microwave sweep while the atoms B remain in $\ket{\downarrow}$, as illustrated in 
	Fig.~\ref{fig:ExpSequ_SI}b. In this procedure, exciting first the atoms in $\ket{\downarrow}$ 
	has the advantage of minimizing the depumping of the $\ket{\uparrow}$ atoms by the addressing 
	light (see Sec.~\ref{SubSM:Decoherence} below). An example of perfect N\'eel configuration obtained 
	at the end of the preparation is shown in Fig.~\ref{fig:ExpSequ_SI}c. 
	
	The experimental sequence (including the detection part detailed in the next Section) 
	is repeated typically over 1000 defect-free assembled arrays. 
	This allows us to calculate the magnetization 
	and the spin correlations by averaging over these realizations.

	\subsubsection{State detection procedure}\label{SubSM:detection}
	
	At the end of the sequence, we read out the state of each atom in the natural $z$-basis. 
	To do so, we deexcite the atoms from $\ket{\uparrow}$ to the $5S_{1/2}$ manifold where 
	they are recaptured in the tweezers and imaged. Thus, the $\ket{\uparrow}$ (resp. $\ket{\downarrow})$ 
	state is mapped to the presence (resp. absence) of the corresponding atom. In order to avoid the 
	detrimental effects of the $\ket{\uparrow}-\ket{\downarrow}$ interaction-induced dynamics 
	during the deexcitation, we freeze out the system by shelving the $\ket{\downarrow}$ atoms 
	to $\ket{D}=\ket{59D_{3/2},m_j = -1/2}$ where they hardly interact with the ones in $\ket{\uparrow}$. 
	This is achieved by using a $48~$ns microwave  $\pi$-pulse at $10.6~$GHz. 
	The subsequent deexcitation is performed by applying a  $2.5~\mu$s light pulse 
	resonant with the transition between $\ket{\uparrow}$ and the short-lived intermediate 
	state $6P_{3/2}$ from which the atoms decay back to $5S_{1/2}$. 
	Additionally, when we want to measure the spins along $x$ we rotate them by 
	applying a $27~$ns microwave $\pi/2$-pulse on the $\ket{\uparrow}-\ket{\downarrow}$ 
	transition prior to the detection. However, this procedure is efficient only for light-shifts 
	$|\delta(t)|$ much smaller than the microwave Rabi frequency, i.e.~for times larger than 
	$\sim0.5~\mu$s during an adiabatic preparation.
	
	\subsection{Experimental imperfections}\label{SM:exp_imperfections}
	
	The sequences described above are affected by experimental imperfections. 
	As taking all of them into account is intractable, we estimate here the effect of the main 
	imperfections on the quantities we measure. We first analyse the state preparation and 
	measurement (SPAM) errors and then discuss decoherence in the system.

	\subsubsection{SPAM errors}\label{SubSM:SPAM_errors}
	
	%%%%%%%%%%%%%%%%% Figure SM 2 %%%%%%%%%%%%%%%%%%%
	\begin{figure*}[t]
		\includegraphics[width = 18.00 cm]{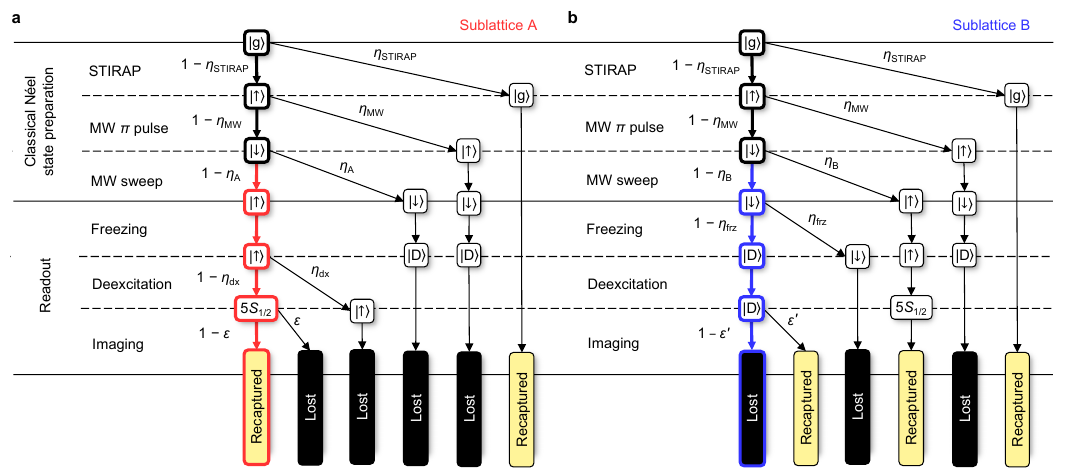}
		\caption{\textbf{Simplified error tree associated to the preparation of the initial N\'eel state,} for {\bf a} 
			the atoms in sublattice A (non-addressed), and {\bf b} in sublattice B (addressed). 
			For simplicity, the events with a probability of order 2 or higher in the $\eta_i$, $\epsilon$, $\epsilon^{'}$ are disregarded. }
		\label{fig:Error_tree}
	\end{figure*}
	%%%%%%%%%%%%%%%%%%%%%%%%%%%%%%%%%%%%%%%%%%%
	
	In order to estimate the SPAM errors, we break down the sequence into a series of steps $i$, each having a small but finite 
	failure probability $\eta_i$. In the following, we keep only the contributions of imperfections to first order in the $\eta_i$'s.  
	
	\renewcommand\arraystretch{1.5}
	\begin{table*}
		\centering
		\begin{tabular}{|c|c|c|c|c|}
			\hline
			\textbf{Stage}& \textbf{Step}&\textbf{Symbol}& \textbf{Value}& \textbf{Main physical origin} \\ \hline \hline
			\multirow{3}{*}{Classical N\'eel state preparation} &
			Rydberg excitation & $\eta_{\rm STIRAP}$ & $5\%$ & \begin{tabular}{ccc}
				Imperfect optical pumping,\\ Laser phase noise,\\ Spontaneous emission 	from $6P_{3/2}$ \cite{Deleseleuc2018}
			\end{tabular}\\ \cline{2-5}
			& MW $\pi$- pulse & $\eta_{\rm MW}$          & $2\%$  &  Effect of $H_{\rm XY}$ during pulse \\ \cline{2-5}
			& MW sweep        & $\eta_{\rm A},\eta_{\rm B}$         & $4\%,5\%$ & Effect of $H_{\rm XY}$ and finite value of $|\delta_0|$ \\ \hline \hline
			\multirow{4}{*}{Readout} & Freezing        & $\eta_{\rm frz}$         & $<1\%$   & Effect of $H_{\rm XY}$ during pulse\\ \cline{2-5}
			& Deexcitation    & $\eta_{\rm dx}$        & $3\%$     & Mechanical effect of deexcitation laser beam\\ \cline{2-5}
			& False $\ket{\downarrow}$  & $\epsilon$        & $1\%$     &  Background gas collisions \cite{Deleseleuc2018}\\ \cline{2-5}
			& False $\ket{\uparrow}$  & $\epsilon^\prime$ & $5\%$ & Rydberg state radiative lifetime  \cite{Deleseleuc2018} \\
			\hline
		\end{tabular}
		\caption{\textbf{Summary of the experimental errors defined in Fig.~\ref{fig:Error_tree}, together with  their main physical origin.}}
		\label{tab:error_tree}
	\end{table*}
	
	As an example, we show in Fig.~\ref{fig:Error_tree} the discretized sequence corresponding to the 
	preparation and measurement of the classical N\'eel configuration (corresponding to the time $t=0$ 
	in Fig.~\ref{fig:fig2}a of the main text). Table~\ref{tab:error_tree} gives the corresponding values of the 
	probabilities $\eta_i$ for 42 atoms, that are either inferred from a series of dedicated experiments, 
	or estimated from numerical simulations. The table also mentions the physical origin of these imperfections. 
	
	For atoms in sublattice A (non-addressed), the error tree leads to the probability to recapture the atoms 
	at the end of the sequence, which reads (to first order):
	
	\begin{equation}
	P_z^{\text{A}}\approx 1 - \eta_{\text{MW}} - \eta_{\text{A}} - \eta_{\text{dx}} - \epsilon  \ .
	\end{equation}
	Similarly, the calculation for sublattice B (addressed atoms) yields:
	\begin{equation}
	P_z^{\text{B}} \approx \eta_{\text{STIRAP}} + \eta_{\text{B}} + \epsilon' \ .
	\end{equation}

	Using the values reported in Table~\ref{tab:error_tree}, we obtain 
	$P_z^{\text{A}} = 0.90$, $P_z^{\text{B}} = 0.15$. 
	From these probabilities, we compute an initial magnetization along 
	$z$, $\sigma_z^{\text{A}} = 2P_z^{\text{A}} - 1 = 0.8$ and 
	$\sigma_z^{\text{B}} = 2P_z^{\text{B}} - 1 = -0.70$. We checked that theses values agree with 
	measured magnetizations at $t=0$, which are used as a calibration of the errors, for both the FM and the AFM (Fig.~\ref{fig:fig2}a,d).
	Finally the error tree allows us to infer the 
	probability of successful initial preparation per spin. We find $0.87$ for the 
	atoms in sublattice A  and $0.92$ for the ones in B. 
	Using the preparation part of the error tree (Fig.~\ref{fig:Error_tree}), we find 
	$1-\eta_{\text{STIRAP}} - \eta_{\text{MW}} - \eta_{\text{A}} = 0.89$ for the atoms 
	in sublattice A  and $1-\eta_{\text{STIRAP}} - \eta_{\text{MW}} - \eta_{\text{B}} = 0.88$ for the ones in B. 
	These values are very similar to the ones including detection errors, indicating that this experiment is dominated by preparation errors.

	\subsubsection{Decoherence during the adiabatic ramp}\label{SubSM:Decoherence}
	
	Besides the SPAM errors described previously, additional imperfections lead to decoherence.
	
	First, we focus on the long-time behaviour of the magnetizations for the $10\times10$ arrays. 
	In Fig.~\ref{fig:fig2}a, one observes that, in the FM case, the $z$-magnetizations 
	of sublattices A and B do not vanish at late times, but reach a constant finite value 
	of a few percent. In contrast, this does not occur in the AFM case (Fig.~\ref{fig:fig2}d). 
	We qualitatively explain this effect by the following observations. 
	First, due to off-resonant scattering by the addressing beam, atoms in 
	$\ket{\uparrow}$ are slowly depumped to the ground state $\ket{g}$; 
	we have measured the effective lifetime of an  addressed $\ket{\uparrow}$ 
	atom to be $\sim 4~\mu$s, whether the light-shift is 
	$2\pi \times15$ or $-2\pi \times15~$MHz (so that this alone, cannot 
	explain the difference between the FM and AFM cases). However, during 
	our adiabatic ramp down of light-shift $\delta(t)$, the addressed atoms are 
	initially in $\ket{\downarrow}$ (and thus cannot be depumped). Depumping 
	sets in only when the system enters the ordered phase, where an addressed 
	atom has a significant probability to be in $\ket{\uparrow}$. 
	Since $\delta_{\rm c}^\text{AFM}<\delta_{\rm c}^\text{FM}$, 
	the addressing beam intensity (and thus the depumping rate) is at this 
	stage much smaller for the AFM case than for the FM case, and thus 
	has a negligible effect in the former case. 
	
\begin{figure}
		\includegraphics[width=8.648cm]{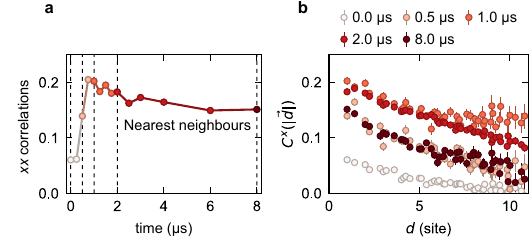}
		\caption{\textbf{Time dependence of the correlations along $x$ in the FM case for a $10 \times 10$ lattice.} 
			\textbf{a}, Time evolution of the nearest-neighbour correlations along $x$ (different colors correspond to different times). 
			\textbf{b}, Spatial correlations as a function of distance, measured at different times 
			$t = \{0.0,~0.5,~1.0,~2.0,~8.0\}~\mu$s indicated by dashed lines in \textbf{a}. }
		\label{fig:SM_correlations}
	\end{figure}

	Second, we investigate the role of decoherence on the appearance of long-range order along 
	$x$ in the FM case, for the $10 \times 10$ array.
	Figure~\ref{fig:SM_correlations}a shows the time evolution of the nearest-neighbour correlations 
	as we ramp down the light-shift, all the way up to 8~$\mu$s (in contrast with Fig.~\ref{fig:fig2}b 
	of the main text where the evolution is shown only up to 3~$\mu$s, and for 42 atoms). 
	Two timescales appear: first, correlations build up until $t\simeq 1~\mu$s as the FM 
	state is adiabatically prepared ; then, they slowly decay and lose $25~$\% of their 
	value in 7~$\mu$s. This decay is not expected, since the system should be ideally 
	in steady state once it has reached the ferromagnetic phase. We conjecture that the 
	experimental system is affected by decoherence arising from a combination of the 
	residual atomic motion and spontaneous emission from the Rydberg states.
	
	To further analyse the evolution of the ferromagnetic order, we probe the full spatial 
	structure of the correlations at different times. Figure~\ref{fig:SM_correlations}b 
	summarizes the results. We observe that for a given distance $d$ all the correlations 
	feature a similar time evolution: a fast increase followed by a slow decay, with a 
	turning point around $1~\mu$s. For this particular point, the data reveal a plateau 
	for distances of more than 6 sites -- the signature of the long range order mentioned 
	in the main text -- that disappears for $t\gtrsim2~\mu$s. 
	This suggests that despite the decoherence present in the system, 
	we are able to observe the long range ordering expected from the dipolar
	interactions over a substantial time window.
	
	%%%%%%%%%%%%%%%% Figure SM 3 %%%%%%%%%%%%%%
	%%%%%%%%%%%%%%%%%%%%%%%%%%%%%%%%%%%%%%%

	\subsection{Ground state properties of the XY model }\label{SM:GS}
	
	We study here the ground states of the Hamiltonians $H_{\rm XY}$ and $H_{\rm XY}+H_{\rm Z}$. 
	We define them as in the main text. First: 
	\begin{align}
	H_{\rm XY} &=  -\frac{J}{2}\sum_{i<j} \frac{a^3}{r_{ij}^3} \left[ \sigma_i^x \sigma_j^x + \sigma_i^y \sigma_j^y \right]\\
	&= -\frac{J}{\hbar^2} \sum_{i<j} \frac{a^3}{r_{ij}^3} \left[ S_i^+ S_j^- + S_j^+S_i^-\right]
	\end{align}
	where $S_i^\pm = S_i^x \pm i S_i^y = \hbar(\sigma_i^x \pm i \sigma_i^y)/2$ are the ladder operators 
	for spin-1/2 degrees of freedom on a square lattice with $N$ sites, $r_{ij}$ is 
	the distance between sites $i,j$ and $a$ is the lattice spacing.
	Second, the on-site Hamiltonian is:
	\begin{equation}\label{eq:Hz}
	H_{\rm Z}  =  \hbar \delta \sum_{i\in B}{\sigma^z_i + 1\over 2}
	\end{equation}
	where the magnitude of the light-shift $\delta$ depends on the intensity of the addressing laser. 
	
	The experimental implementation has a ferromagnetic coupling, 
	$J/h = 0.77$ MHz, and to study antiferromagnetism one must prepare negative temperature states.
	For theoretical purposes, however, we treat $J$ as a free parameter and frame the discussion in 
	terms of the ground state physics of $H_{\rm XY}$ with either ferromagnetic ($J>0$) 
	or antiferromagnetic ($J<0$) coupling. 
	We refer to them as $H_{\rm XY}^{\rm FM}$ and $H_{\rm XY}^{\rm AFM}$. 
	
	It is natural to compare the dipolar $H_{\rm XY}$ to the nearest-neighbor XY model on the square lattice, 
	\begin{equation}\label{eq:Hnn}
	H_{\rm nn} = - \frac{J}{2} \sum_{\langle i j \rangle} \sigma^x_i\sigma^x_j + \sigma^y_i \sigma^y_j,
	\end{equation}
	where $\langle i j \rangle$ are pairs of neighboring sites, with $i<j$. 
	For $H_{\rm nn} $, the sign of the coupling $J$ is unimportant, 
	as $U_A H_{\rm nn} U_A^\dagger = -H_{\rm nn} $, with $U_A = \prod_{j\in A} e^{-i \pi S_j^z}$. 
	In 1988, Kennedy, Lieb, and Shastry rigorously proved that the unique ground state of $H_{\rm nn} $
	has long-range XY order (LRO)~\cite{kennedyXYModelHas1988}.

	For models with long-range interactions, there are analogous mathematical theorems 
	for classical systems at finite temperature, and for quantum systems in which the interaction 
	strength depends on the Manhattan distance $\|r_i - r_j\|_1$~\cite{frohlichPhaseTransitionsReflection1978}. 
	In a recent work, Bj\"ornberg and Ueltschi
	addressed quantum spin-$S$ models with  
	interactions depending on the Euclidean distance $\|r_i-r_j\|_2$, 
	although their results require large $S$ and spatial dimension three or higher \cite{bjornbergReflectionPositivityInfrared2022a}. 
	Absent a rigorous proof of LRO for the two-dimensional, spin-$1/2$, dipolar XY model, 
	one can study it using semi-analytic spin wave theory and various numerical 
	methods~\cite{maleevDipoleForcesTwodimensional1976, Peter2012, yaoQuantumDipolarSpin2018}. 
	In a companion paper \cite{QuantumPhaseDiagramInPreparation}, we investigate the ground-states of
	$H_{\rm XY}$ on various geometries, such as tori and infinite cylinders, 
	with an eye towards the thermodynamic limit, $N\to \infty$.
	Here, we restrict our focus to finite rectangular arrays as probed in the experiment, and use 
	$H_{\rm nn} $ as a reliable benchmark for comparison.

	\subsubsection{Symmetries, magnetization sectors, and order}\label{SubSM:Symmetry}
	
	As emphasized in the main text, $H_{\rm XY}$ possesses the continuous symmetry:
	$U_z(\theta) H_{\rm XY} U_z(-\theta) =  H_{\rm XY}$ with
	\begin{equation}
	U_z(\theta) = \exp(-i  \sum_j \theta S_j^z/\hbar) =\exp(-i \theta M^z / 2)
	\end{equation}
	This operator is generated by the total magnetization, $M^z = \sum_i \sigma_i^z$, 
	and represents the Lie group $U(1) \cong SO(2)$.  
	Additionally, $H_{\rm XY}$ is invariant under the $\mathbb{Z}_2$ Ising symmetry, 
	$\alpha_{2} : (\sigma^x, \sigma^y, \sigma^z) \to (\sigma^x, -\sigma^y, -\sigma^z)$,
	as well as any spatial symmetries of the lattice, such as translation or rotation. 
	This model is also time-reversal-symmetric, as represented by the anti-unitary operator 
	$\mathcal{T} = \mathcal{C}$, where $\mathcal{C}$ applies complex conjugation. 
	Here $\mathcal{T}$ differs from the usual $SU(2)$ time-reversal symmetry, 
	which applies the unitary spin rotation $U_{y}(\pi) = \exp(- i \pi M^y/2)$ in addition to $\mathcal{C}$.
	Our atypical choice of $\mathcal{T} = \mathcal{C}$ allows it to remain a symmetry 
	in the presence of the on-site perturbation, $H_{\rm Z}$.
	
	In a finite, closed quantum system, all eigenstates $\ket{\psi_n}$ of $H_{\rm XY}$ 
	can be chosen to be simultaneous eigenstates of all of these symmetry operators.  
	In particular, they are eigenstates of the total magnetization, $M^z \ket{\psi_n} =  \lambda^z_n \ket{\psi_n}$, 
	and so can be collected into magnetization sectors, conventionally labeled by $S^z = M^z/2$.
	As a consequence, all $M^z$-non-conserving operators such as $\sigma_i^x$ and $\sigma_i^y$ 
	have identically vanishing expectation values, 
	$\langle \sigma_i^x \rangle = \langle \sigma_i^y \rangle = 0$,
	in any energy eigenstate $\ket{\psi_n}$, or in any superposition of eigenstates 
	within the same magnetization sector.

	In the experiment, systematic errors in the measurement process lead to a small, 
	nonzero $\langle \sigma_i^x \rangle \ne 0$. 
	This value is not a consequence of the physics we are interested in.
	When analyzing the experimental data, we thus choose to nullify any single-spin 
	contributions by using the \textit{connected} correlator,
	\begin{equation}
	C^x(i,j) = \langle \sigma_i^x \sigma_j^x \rangle  - \langle \sigma_i^x \rangle \langle \sigma_j^x \rangle
	\end{equation}
	In the special case of $M^z$ eigenstates with $\langle \sigma^x \rangle = 0$, 
	$C^x(i,j) = \langle \sigma_i^x \sigma_j^x\rangle$. 
	This correlation function is not generically zero. If $|C^x(i,j)|$ approaches a constant $C^x_\infty>0$ 
	for distantly separated spins $i,j$, then the corresponding state is said to possess \textit{long-range XY order} 
	or \textit{off-diagonal long-range order} (LRO)~\cite{yangConceptOffDiagonalLongRange1962}. 
	Such LRO is the defining feature of continuous symmetry breaking in finite quantum systems.

	Rather than the long-distance plateau, an equally good order parameter for $U(1)$ 
	symmetry breaking is given by the in-plane magnetization squared
	\begin{equation}\label{eq:mx2}
	m_{\rm FM/AFM}^2 = \frac{1}{N^2}\sum_{i,j} (\pm 1)^{(r_i^x + r_i^y)/a}C^x(i,j)
	\end{equation}
	where $a$ is the lattice spacing, and the sign is taken $+1$ for $m_{\rm FM}^2$, and $-1$ for $m_{\rm AFM}^2$.
	In the ${N\to\infty}$ limit, any state with a correlation plateau $C^x_\infty\ne 0$ 
	will also have a finite magnetization $m_{\rm FM/AFM}^2$, and vice versa~\cite{tasakiLongRangeOrderTower2019}.

	When continuous symmetry breaking occurs in the thermodynamic limit, then at finite size the 
	lowest energy state in each $S^z$ sector will be approximately,
	\begin{equation}
	    \ket{\Gamma_s^{\mathrm{FM/AFM}}} = \frac{1}{\mathcal{N}_s} \int_0^{2\pi}\frac{\dd\theta}{2\pi}  e^{is \theta} \ket{\theta^{\mathrm{FM/AFM}}}
	\end{equation}
	where $\ket{\theta^{\mathrm{FM/AFM}}}$ is the classical, symmetry-breaking product state where each spin points at angle $\theta$ or $-\theta$ 
	in the $xy$-plane, $s$ is an integer specifying the $S^z$ sector, and $\mathcal{N}_s$ is a normalization factor. 
	Known either as the Anderson tower or Dicke states, $\ket{\Gamma_s}$ are angular momentum eigenstates of an emergent rigid rotor degree of 
	freedom describing the collective orientation of all the spins in the system~\cite{andersonApproximateQuantumTheory1952, andersonBasicNotionsCondensed2010, 
		tasakiLongRangeOrderTower2019, beekmanIntroductionSpontaneousSymmetry2019a}. 
	The true ground states in each $S^z$ sector are also dressed by quantum spin wave fluctuations, which weaken the magnetic order~\cite{andersonApproximateQuantumTheory1952}.
	For the ideal case of a uniform superposition over fully spin-polarized states  $\ket{\theta^{\mathrm{FM/AFM}}}$, the correlations in $\ket{\Gamma_0}$ lead to $C^x_\infty = m^2 =0.5$, plus $1/N$ corrections. 
	The effective in-plane magnetization of a $U(1)$-symmetric state should thus be identified as $m_{\mathrm{eff}} \equiv \sqrt{2 C^x_\infty}$. 
	That is, if one were to add a small symmetry-breaking field, then the corresponding non-symmetric ground state would have an average magnetization $\langle \sigma^x \rangle = m_{\mathrm{eff}}$.

	\subsubsection{DMRG calculations}\label{SubSM:DMRG}
	
	%%%%%%%%%%%%%%%% Figure SM 5 %%%%%%%%%%%%%%%%
	\begin{figure*}
		\centering
		\includegraphics{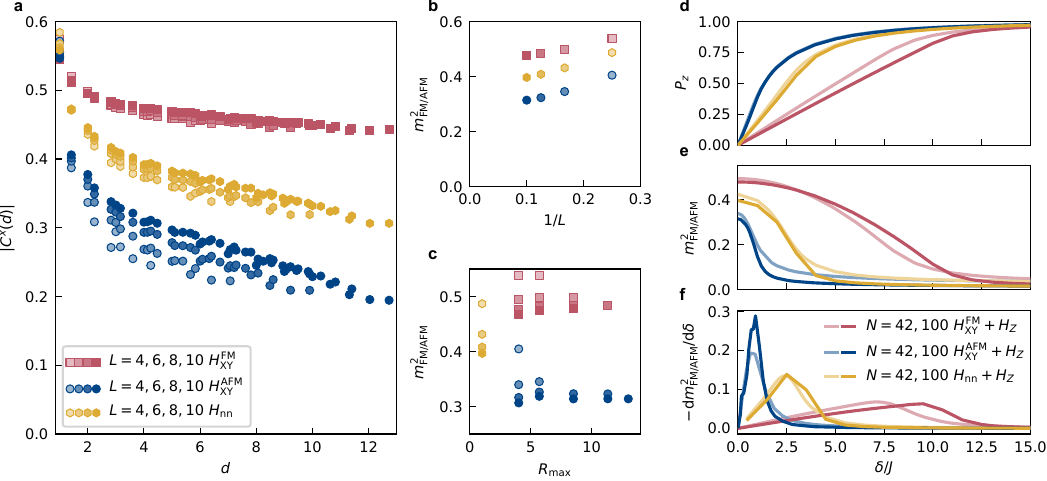}
		\caption{\textbf{DMRG ground state calculations.}
			\textbf{a}, Real-space correlation profile $|C^x(d)|$ on $L\times L$ 
			square clusters with open boundary conditions. The ground state of $H_{\rm XY}^{\rm FM}$ clearly exhibits XY LRO at 
			all system sizes. For $H_{\rm XY}^{\rm AFM}$ and $H_{\rm nn}$, the correlations decrease at long distances, 
			but this decay is reduced as $L$ increases. 
			\textbf{b}, Finite-size scaling of the magnetization $m_{\rm FM/AFM}^2$. All three models are consistent with 
			$m_{\rm FM/AFM}^2 > 0$ as $L\to \infty$. 
			\textbf{c}, Dependence of $m_{\rm FM/AFM}^2$ on the interaction distance cutoff $R_{\rm max}$. 
			At each system size, the ground state correlations are well-converged by $R_{\rm max}\approx 4$.  \textbf{d-f}, 
			Ground state properties of $H_{\rm XY} +H_{\rm Z}$ as a function of $\delta$.  
			There is a smooth crossover from the XY ordered state at $\delta=0$ to the staggered paramagnet 
			as $\delta \to \infty$.  The $-\dd m^2/ \dd \delta$ peaks ($\textbf{f}$) are finite-size incarnations of the 
			quantum phase transition expected in the thermodynamic limit; we use their centers to define the crossover point $\hbar \delta_c/J$. }
		\label{fig:ground_dmrg}
	\end{figure*}
	%%%%%%%%%%%%%%%%%%%%%%%%%%%%%%%%%%%%%%%%%

	For a numerical investigation of the ground states, we apply the density 
	matrix renormalization group (DMRG) algorithm~\cite{whiteDensityMatrixFormulation1992}. 
	We employ the general matrix product state (MPS) framework implemented in the  TeNPy 
	software library~\cite{hauschildEfficientNumericalSimulations2018}.  
	While MPS are best-representative of one-dimensional quantum systems, it is now routine to apply 
	DMRG to two-dimensional models under certain geometric 
	restrictions~\cite{stoudenmireStudyingTwoDimensionalSystems2012}. 
	We always work with charge-conserving tensors that respect the $U(1)$ symmetry of the Hamiltonian. 
	
	To begin, we use DMRG to compute the ground state of $H_{\rm XY}$ and $H_{\rm nn} $ 
	on $L\times L$ square clusters with open boundary conditions, for $L=4,6,8,$ and $10$.  
	With all-to-all interactions included, we reliably obtain well-converged states at relatively low 
	MPS bond dimensions, $\chi$, as quantified by the truncation error of the discarded 
	Schmidt states, $\epsilon_{\rm trunc}$.
	The most difficult finite system we study is $H_{\rm XY}^{\rm AFM}$ on the $10\times 10$ lattice, 
	for which $\epsilon_{\rm trunc}<10^{-5}$ at $\chi=2048$. All other cases achieve the same or 
	better convergence by $\chi=1024$, or even $\chi<200$ on the smaller systems.

	All DMRG ground states feature the strong $\langle{\sigma^x\sigma^x}\rangle$ 
	correlations expected in an XY LRO state. 
	In Fig.\,\ref{fig:ground_dmrg}a, we show the real-space correlation profile $C^x(d)$, 
	which averages $C^x(i,j) = \langle \sigma^x_i \sigma^x_j \rangle$ over all spins $i,j$ 
	separated by a displacement vector $\vec{d}_{ij}$ with length $d$.  
	The long-range-interacting ferromagnet, $H_{\rm XY}^{\rm FM}$, exhibits a clear plateau in $C^x(d)$ 
	at long distances for all system sizes. 
	Such a plateau is less apparent for $H_{\rm XY}^{\rm AFM}$ and $H_{\rm nn} $, although for either 
	model $C^x(d)$ is still quite large at the longest distances.  
	Furthermore,  $C^x(d)$ increases with $L$ in both models, suggesting the spatial decay 
	of $C^x(d)$ is amplified by finite-size effects. 
	
	We also look for a finite squared magnetization, $m_{\rm FM/AFM}^2$. 
	We plot the finite-size dependence of this quantity in Fig.\,\ref{fig:ground_dmrg}b, 
	which is consistent with $m_{\rm FM/AFM}^2>0$ as $L\to \infty$.
	To further test the effects of the long-range interactions, we introduce a cutoff radius $R_{\rm max}$, 
	and only include interactions between spins $i,j$ separated by distance $d_{ij} < R_{\rm max}$.  
	We find that ground state properties converge quickly with respect to this approximation parameter; 
	the long-range interactions do not induce a quantum phase transition in either model.
	In Fig. \ref{fig:ground_dmrg}c, we show the dependence of $m_{\rm FM/AFM}^2$ on $R_{\rm max}$, finding that, 
	at fixed system size, it is not strongly dependent on $R_{\rm max}>4$.
	This is not too surprising: with the moderately fast $1/r^3$ decay, the interaction strength beyond this 
	point is on the order of $0.01$ $J$ or less.
	
	Overall, $H_{\rm XY}^{\rm FM}$ is clearly XY LRO, while $H_{\rm XY}^{\rm AFM}$ and $H_{\rm nn} $ 
	exhibit stronger finite-size effects. 
	Given that $H_{\rm nn} $ is rigorously known to be LRO in the thermodynamic limit, 
	the similar behavior observed for $H_{\rm XY}^{\rm AFM}$ is a strong indication that it is as well. 
	
	\subsubsection{Quantum phase diagram of \texorpdfstring{$H_{\rm XY}+H_{\rm Z}$}{HXY+HZ}} \label{SubSM:QPT}
	
	We now investigate the ground state phase diagram in the presence of the externally applied light-shift $\delta$, described by
	$H_{\rm Z}$ (Eq.\,\ref{eq:Hz}).
	This perturbation preserves the $U(1)$ symmetry of $H_{\rm XY}$, as well as the anti-unitary time-reversal symmetry.
	On the other hand, $H_{\rm Z}$ breaks the Ising symmetry $\sigma^z_i \to -\sigma^z_i$, 
	and reduces the spatial rotation and translation symmetries.
	For sufficiently large $\delta$, the lowest energy state of $H_{\rm XY}+H_{\rm Z}$ has $M^z\ne 0$, 
	but such states are dynamically decoupled from the $S^z=0$ sector in which the adiabatic preparation protocol takes place. 
	Henceforth, we always consider the ground states within the $S^z=0$ sector, 
	as these are the ones most relevant to the experiment.
	
	Because the perturbation $H_{\rm Z}$ is $U(1)$ symmetric, 
	the XY LRO phase of $H_{\rm XY}$ may be stable to a sufficiently small staggered field.  
	Microscopically, the dominant effect of a small $\delta$ should be to slightly cant the spins towards the $z$-axis. 
	This will in turn modify the spin stiffness and the spin wave velocity, but not destroy the underlying order.
	By contrast, when $\delta$ is very large, the ground state must be a gapped, trivial paramagnet, 
	in which $\langle \sigma^x \sigma^x\rangle$ correlations decay to zero at long 
	distances~\cite{hastingsSpectralGapExponential2006}.  
	Between these two limits, we expect a quantum phase transition (QPT) at some critical value, 
	$\delta_c$, of the applied field. 
	In a companion paper \cite{QuantumPhaseDiagramInPreparation}, 
	we investigate this QPT in detail, finding that, in the thermodynamic limit,
	it is likely a continuous, second-order transition. 
	For $H_{\rm nn} +H_{\rm Z}$, the transition is in the 3D XY universality class.
	For the $1/r^3$ models, the standard theory expectation is that the AFM QPT is in the same universality class as the short-range model (i.e. 3D XY), 
	while the FM QPT is in a different universality class with mean-field-like critical exponents~\cite{Defenu2021}.
	
	Here, we focus our attention on the  $6 \times 7$ and $10 \times 10$ arrays studied in the experiment.
	We calculate the $S^z=0$ ground state of $H_{\rm XY} + H_{\rm Z}$ at various 
	light-shifts $\delta$ using DMRG. 
	At these system sizes, the sharp QPT expected in the thermodynamic limit is smoothed to a 
	broad crossover between the XY-ordered phase for small $\delta$, and a trivial paramagnet for large $\delta$.
	Three features of this crossover are shown in Fig. \ref{fig:ground_dmrg}d-f.  
	
	First, in Fig. \ref{fig:ground_dmrg}d, we plot the staggered $\sigma^z$ polarization,
	\begin{equation}
	P_z = \frac{1}{N} \sum_{i\in A} \langle \sigma_i^z \rangle - \frac{1}{N} \sum_{i\in B} \langle \sigma_i^z \rangle 
	\end{equation}
	which measures the alignment with the staggered field $H_{Z}$.
	For large $\delta \gg \delta_c$, the ground state approaches the staggered product state used 
	to initialize the adiabatic ramp in the experiment, and the polarization saturates to $P_z = 1$.
	For $\delta =0$, $P_z=0$ due to the Ising symmetry of $H_{\rm XY}$, which enforces 
	$\langle \sigma_i^z \rangle =0$.
	We emphasize  that $P_z = 0$ is \textit{not} a generic feature of the XY-ordered phase. 
	Indeed, for small $\delta < \delta_c$, the spins partially align with the applied field, yielding $P_z > 0$. 
	
	Figure \ref{fig:ground_dmrg}e displays the complementary behavior for the magnetization, 
	$m_{\rm FM/AFM}^2$.
	At small $\delta$, the field-induced canting of the spins towards the $z$-axis causes 
	$m_{\rm FM/AFM}^2$ to decrease proportionally to $\delta^2$.
	At large $\delta$, the ground state approaches the (staggered) $z$-aligned product state, 
	in which $m_{\rm FM/AFM}^2 =0$. 
	The magnetization changes most rapidly at the crossover, giving rise to the clear peaks in 
	$\dd m_{\rm FM/AFM}^2 / \dd \delta$ shown in Fig.\,\ref{fig:ground_dmrg}f. 
	We take the center of these peaks as our definition of the crossover point, $\delta_c^{\rm FM/AFM}$. 
	For the $N=42$ cluster, the values are $\hbar \delta_c^{\rm FM}/J = 7.1(3)$, $\hbar \delta_c^{\rm AFM}/J = 0.8(1)$, 
	and $\hbar \delta_c^{\rm nn}/J = 2.4(1)$.  For the $N=100$ cluster, we find $\hbar \delta_c^{\rm  FM}/J = 9.5(3)$,  
	$\hbar\delta_c^{\rm AFM}/J = 0.9(1)$, and $\hbar \delta_c^{\rm nn}/J = 2.5(9)$.
	As $N\to \infty$, the smooth crossover is expected to sharpen into a \textit{bona fide} QPT, 
	and $m_{\rm FM/AFM}^2(\delta)$ will be non-analytic at the critical point.

	\subsection{Adiabatic preparation - theory and numerics}\label{SM:Adiabatic}
	
	We now provide theoretical and numerical analyses of the adiabatic preparation protocol used in the experiment. 
	As mentioned above, we study both the FM and AFM cases considering 
	$H_{\rm XY}^{\rm AFM}= - H_{\rm XY}^{\rm FM}$.
	Additionally, for a time-reversal-symmetric Hamiltonian such as $H=H_{\rm XY} + H_{Z}$, 
	the dynamics under $H(t)$ and $-H(t)$ are identical (as long as the initial state is also 
	time-reversal-symmetric)~\cite{sorensenAdiabaticPreparationManybody2010}.
	So for a finite-time (quasi-adiabatic) ramp, the diabatic errors incurred attempting to 
	follow the topmost state of $H_{\rm XY}^{\rm FM} + H_{Z}$ are the same as for a  
	ground-state protocol with $H(t) = H_{\rm XY}^{\rm AFM} - H_{Z}(t)$.
	
	\subsubsection{Excitation gaps and an alternative protocol}\label{SubSM:adiabatic_alternative}
	
	%%%%%%%%%%%%%% Figure SM 6 %%%%%%%%%%%%%%%%%
	\begin{figure}[t]
		\centering
		\includegraphics[width=\columnwidth]{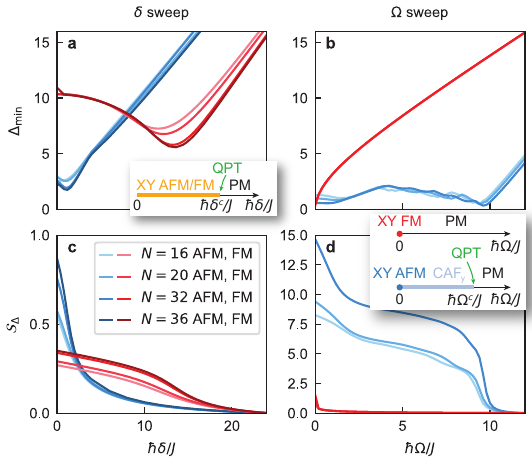}
		\caption{\textbf{Excitation gap for two adiabatic preparation protocols. }
			\textbf{a}, Minimal energy gaps of $H_{\rm XY}^{\rm AFM (FM)}+H_{\rm Z}$ to the lowest 
			excited state in the $M^z=0$ sector as a function of $\hbar\delta/J$. 
			We here only consider gaps among states with momentum $\vec{k}=0$ and fully symmetric under 
			the lattice point-group, which reflects the setup in the (ideal) experiment.
			Blue (red) curves show the results for the AFM (FM) model. Darker colors correspond to larger system sizes.
			The inset shows a sketch of the expected phase diagram.
			\textbf{b}, Same as \textbf{a}, but for the protocol with Hamiltonian $H_{\rm XY}^{\rm AFM (FM)} + H_{\rm X}$. 
			Here we cannot restrict the analysis to a single $M^z$ sector since it is not conserved.
			\textbf{c}, Cumulatively integrated $1/\Delta_{\rm min}^2$ [starting from the largest value
			$\hbar\delta/J = 24$] for the gaps shown in \textbf{a}. The values at $\hbar\delta/J = 0$ 
			measure how difficult it is to prepare the ground state of $H_{\rm XY}^{\rm AFM  (FM)}$ by sweeping $\delta$.
			\textbf{d}, Same as \textbf{c} but for the gaps along $\Omega$, as shown in \textbf{b}.
			The inset shows a sketch of the expected phase diagrams for $H_{\rm XY}^{\rm AFM (FM)}+H_{\rm X}$. }
		\label{fig:Gaps_Protocols}
	\end{figure}
	%%%%%%%%%%%%%%%%%%%%%%%%%%%%%%%%%%%%%

	The success of any finite-duration adiabatic protocol depends crucially on the low-energy spectrum 
	of the system.
	In particular, as the smallest excitation gap encountered along the chosen path through parameter space decreases, the time required to obtain a final, 
	high-fidelity ground state increases.
	To this end, we computed the minimal energy gaps, $\Delta_{\rm min}$, using exact diagonalization 
	on finite clusters with periodic boundary conditions.  
	
	In Fig.\,\ref{fig:Gaps_Protocols}a, we plot the instantaneous gap $\Delta_{\rm min}$ of 
	$H_{\rm XY}^{\rm FM/AFM} + H_{\rm Z}$, in the $S^z=0$ sector, as a function of the light-shift $\hbar \delta/J$.
	We expect the gap for either case to be smallest near the quantum phase transition (Methods~\ref{SubSM:QPT}): 
	for the FM, this dip is seen at $\hbar \delta /J \approx 12$, while in the AFM the gap is minimal when $\hbar \delta/J \lesssim 2$. 
	The size of the minimal gaps decreases with increasing system size $N$ (darker colors), as one would expect at a QPT.
	However, we find the minimal finite-size gaps for the FM model are always larger than the ones for the AFM model. 
	This indicates that for the dipolar XY model,
	the FM requires less total ramp time to prepare than the AFM.

	Besides the staggered light-shift ramp demonstrated in the main text, one can conceive a different route for 
	preparing XY-ordered states: tune down a spatially \textit{uniform} field in the $x$ direction from 
	large values $\hbar \Omega \gg J$ to zero.
	This is similar to what is done in Rydberg quantum simulations of the two-dimensional Ising 
	model~\cite{ Lienhard2018,Scholl2021,Ebadi2021}, and was used in a prior 
	experiment to prepare the topological ground state of a one-dimensional XY 
	model~\cite{deleseleucObservationSymmetryprotectedTopological2019}.
	The corresponding Hamiltonian is 
	$H_{\rm XY}^{\rm AFM (FM)} + H_{\rm X}(t)$, with $H_{\rm X}(t) = \hbar\Omega(t) \sum_i \sigma_i^x / 2$.
	Note that $M^z$ is no longer conserved in the presence of $H_{\rm X}$.
	
	Figure~\ref{fig:Gaps_Protocols}b shows the smallest energy gap 
	for this alternative protocol. 
	The behaviour is very different from the one for the $\delta$ sweep discussed above. 
	For FM interactions, the gap does not show any local minimum and remains large until the end of the sweep, 
	where it finally narrows.
	By contrast,  the gap for the XY AFM is small in  the whole region ${\hbar\Omega/J \lesssim 10}$.
	Based on previous studies of the nearest-neighbor XY model~\cite{Jensen2006, Kar2017}, both of these results are 
	likely a consequence of the expected phase diagram for $H_{\rm XY}^{\rm FM/AFM}+H_{\rm X}$, 
	which we sketch in the inset of Fig.\,\ref{fig:Gaps_Protocols}d.
	For the XY FM, $H_{\rm X}$ is a relevant perturbation to the ordered phase: any 
	non-zero $\Omega$ breaks the $U(1)$ symmetry and, in the thermodynamic limit, immediately destroys the LRO, 
	resulting in a paramagnetic (PM) phase.
	The AFM is also XY-ordered only at the $U(1)$-symmetric point $\Omega=0$, but a small $\Omega$ instead ``cants'' the 
	AFM order towards the $y$-direction by a 
	spin-flop process~\cite{Jensen2006,Kar2017}. 
	The ground state is then still an antiferromagnet, but one ordered along the $y$-direction, i.e. it spontaneously breaks
	the remaining $\mathbb{Z}_2$ symmetry $\sigma^y \to -\sigma^y$ of $H_{\rm XY}+H_{\rm X}$. 
	This ``canted'' antiferromagnet (CAF$_\text{y}$) is stable up to a critical value $\hbar\Omega_c/J$ 
	where it finally undergoes a $2+1D$ Ising QPT to the PM phase~\cite{Kar2017}.
	
	Comparing the gap landscapes in Fig.~\ref{fig:Gaps_Protocols}a,b suggests 
	that preparing the XY AFM requires less time when using $\delta$ sweeps instead of the $\Omega$ sweeps.
	To quantify this, we integrate the squared inverse gaps and define
	\begin{equation}
	\mathcal{S}_{\Delta}(\lambda) = \int_{\lambda_0}^{\lambda} {1\over \Delta_{\rm min}(\lambda')^2} \, \dd \lambda'
	\end{equation}
	where $\lambda =\hbar\delta/J$ or $\hbar\Omega/J$ is the dimensionless parameter for either protocol.
	As one motivation for this quantity, we consider the fidelity susceptibility, $\chi_F$, which is the leading term in the expansion of the fidelity 
	$F(\lambda, \lambda+\delta\lambda) = \left| \langle \psi_0(\lambda) | \psi_0(\lambda+\delta\lambda) \rangle \right|$ of the ground 
	states $\ket{\psi_0(\lambda)}$ between two close points $\lambda$ and $\lambda+\delta\lambda$ in parameter space~\cite{Gu2010}, 
	\begin{equation}
	F(\lambda+\delta\lambda) = 1 - \frac{\delta \lambda^2}{2} \chi_F + \dots
	\end{equation}
	The coefficient $\chi_F$ characterizes how quickly the ground state changes with $\lambda$.
	For a ramp protocol of the form $H(\lambda) = H_{XY} + \lambda  H_{I}$,
	one can show
	\begin{equation}
	\chi_F = \sum_{n \ne 0 } \frac{|\langle \psi_n(\lambda)|H_I| \psi_n(\lambda)\rangle |}{(E_n(\lambda)- E_0(\lambda))^2}
	\end{equation}
	where $\ket{\psi_n(\lambda)}$ is the $n$-th eigenstate of $H(\lambda)$ and $E_n(\lambda)$ is the corresponding energy~\cite{Gu2010}. 
	If we assume that the $n=1$ term is dominant, and the numerator is nearly constant, we get the relationship
	$\chi_F \sim 1/(E_1(\lambda)-E_0(\lambda))^2 = 1/ \Delta_{\rm min}(\lambda)^2$.
	The integral $\mathcal{S}_{\Delta}$ therefore
	estimates the total difficulty of  adiabatically preparing
	the ground state of $H(\lambda)$, starting from the ground state of $H(\lambda_0)$.
	
	In Fig.\,\ref{fig:Gaps_Protocols}c,d, we plot $\mathcal{S}_{\Delta}(\lambda)$ for the two protocols.
	The initial point $\lambda_0$ is taken to be in the paramagnetic phase: $\lambda_0 = 12$ for the $\delta$ 
	sweep and $\lambda_0 = 24$ for the $\Omega$ sweep.
	In either case, $\mathcal{S}_\Delta$ for the AFM (blue curve) exceeds that of the FM as $\lambda \to 0$, 
	indicating that the AFM is more difficult to prepare. 
	Most importantly, comparing Fig.\,\ref{fig:Gaps_Protocols}c,d, one sees that the $H_Z(t)$ protocol is much 
	more efficient at preparing the XY ordered state ($\lambda = 0$) than the $H_X(t)$ protocol, especially for the AFM.

	\subsubsection{Time-dependent MPO-MPS simulation}\label{SubSM:adiabatic_setup}
	
	To ensure that we have a good understanding of the experiment and its imperfections, 
	we also perform numerical simulations of the full many-body quantum dynamics for the $N=42$ adiabatic ramp. 
	We simulate the dynamics in the spin-1/2 subspace, taking into account the error tree in  
	Fig.~\ref{fig:Error_tree} by sampling the state preparation errors with $N_{\rm dis} = 20$ independent simulations.
	
	Atoms that were not excited in the STIRAP with $\eta_\mathrm{STIRAP}=0.03$ correspond 
	to missing sites in the square lattice not taking part in the dynamics. On the remaining sites, 
	we prepare an initial MPS as product state, flipping individual spins according to the probabilities 
	of the microwave $\pi$-pulse, $\eta_\mathrm{MW}=0.003$, and the subsequent microwave sweep 
	of the addressed atoms, $\eta_\mathrm{A}=0.10, \eta_\mathrm{B}=0.03$.
	These values are slightly different from those reported in Table \ref{tab:error_tree}, 
	reflecting an earlier calibration of the experiment. 
	We further update the atom distances $r_{ij}$ in $H_{\rm XY}$ to account for positional disorder: 
	we first take a normal-distributed initial displacement 
	from the square lattice with variance $\sigma_r  =0.2\,\mu\mathrm{m}$, followed by a movement 
	during the dynamics with normal-distributed (time-independent) velocity of variance 
	$\sigma_v = 0.05\, \mu\mathrm{m} / \mu\mathrm{s}$ corresponding to the temperature of the atoms. 
	
	We then time-evolve the states under the time-dependent Hamiltonian, 
	\begin{multline}
	H(t) = - J \sum_{i<j}\frac{a^3}{r_{ij}^3(t)} \left[ S_i^+ S_j^ - + S_i^- S_j^+\right] + H_{\mathrm{vdW}} \\
	+\delta(t) \epsilon_{\rm AFM} \sum_{i\in B} \frac{1+\sigma_i^z}{2} 
	\end{multline}
	where $J/h = 0.77$ MHz, $\delta(t)$ is the ramp shown in Fig.~\ref{fig:mps_exp_comparison}a,c (insets), and $\epsilon_{\rm AFM}= -1$ 
	for the antiferromagnet ($+1$ for the ferromagnet).
	The additional term, $H_{\mathrm{vdW}}$, accounts for the van der Waals interactions between 
	the Rydberg atoms, and takes the form
	\begin{multline}
	H_{\mathrm{vdW}} = \sum_{i<j} \frac{a^6}{r_{ij}^6(t)}\big[U_{6}^{PP} P_i^\uparrow P_j^\uparrow  + 
	U_{6}^{SS} P_i^\downarrow P_j^\downarrow  \\ + U_{6}^{SP} (P_i^\uparrow P_j^\downarrow + P_i^\downarrow P_j^\uparrow)  \big] 
	\end{multline}
	where $P_{i}^{\uparrow/\downarrow} = S_i^z \pm 1/2$ are single-spin projectors. 
	The values of the $U_6$ coefficients are $U_6^{PP}/h = -0.008$ MHz, 
	$U_{6}^{SS}/h = 0.037$ MHz, and $U_6^{SP}/h = -0.0007$ MHz.
	For the purposes of this simulation, we restrict the interaction range of $H_{\rm XY}$ 
	and $H_{\rm vdW}$ to $R_{\rm max} < 3.7$. 
	\textit{}
	We use the $W_{II}$ method \cite{Zaletel2015TimeEvolving} to approximate the 
	evolution operator $e^{-i(H / \hbar) \mathrm{d}t}$ as a matrix product operator (MPO), 
	in combination with standard variational MPO-MPS compression methods.
	Our scheme is correct to first order in the time step $\mathrm{d}t=0.01 \,\mu \mathrm{s}/2\pi$.
	Since the evolution is sufficiently adiabatic, a moderate bond dimension of $\chi=128$ 
	is enough to capture the correlations.  
	In the DMRG ground state, the truncation error at this bond dimension is $6 \times 10^{-7}$ 
	for the ferromagnet, and $3\times 10^{-5}$ for the antiferromagnet. 
	
	When evaluating expectation values and correlation functions from the time-evolved MPS ($t$-MPS), 
	we further account for the measurement errors 
	$\eta_\mathrm{frz}=0.01, \eta_\mathrm{dx}=0.03, \epsilon=0.01, \epsilon'=0.07$ of the error tree. 
	This can be done exactly (without another sampling procedure), 
	since the MPS gives full access to the probabilities of the individual measurement outcomes. 
	
	There are two notable experimental imperfections that we do not take into account in these simulations. 
	First, there are further sources of decoherence in the experiment as discussed in~\ref{SubSM:Decoherence}.
	Second, in our numerical simulations, we assume that all errors in the error tree occur 
	independently for each atom and result in an initial product state of up or down spins or vacant holes.
	Yet, the STIRAP and microwave pulses leave the atoms in coherent superpositions of the relevant atom levels. 
	
	\subsubsection{Simulation results for  \texorpdfstring{$N=42$}{N=42}}\label{SubSM:mps_results}
	
	%%%%%%%%%%%%%% Figure SM 7 %%%%%%%%%%%%%%%%%
	\begin{figure*}[t]
		\centering
		\includegraphics{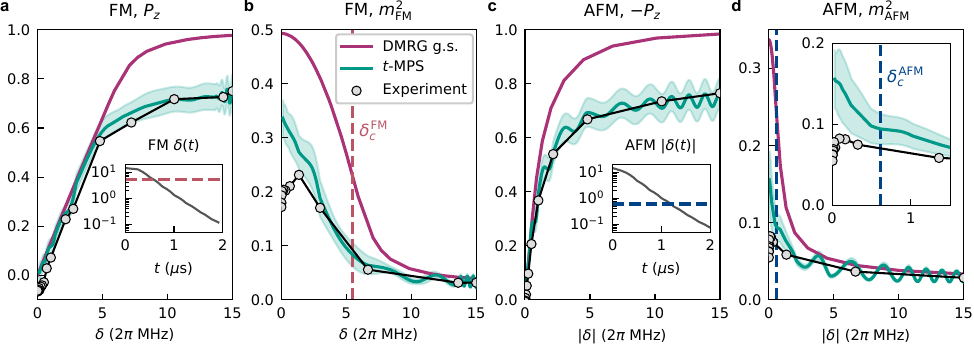}
		\caption{\textbf{Numerical simulation of the adiabatic preparation for the $6 \times 7$ lattice.} 
			We compare the predictions from the  $t$-MPS simulations (disorder ensemble average in dark teal, 
			standard deviation in light teal) to the experimental data (gray), as measured at light-shift $\delta(t)=\delta$. 
			We also show the ground-state expectation value from DMRG (purple)
			\textbf{a}, 
			The staggered polarization $P_z$ of the FM. 
			Theory and experiment agree remarkably well, except for an offset at small $\delta$, due to the light-shift-induced depumping. 
			Inset: ramp $\delta(t)$ used for the FM simulation.
			\textbf{b},
			The ferromagnetic magnetization $m_{\rm FM}^2(\delta)$.  We find excellent agreement between experiment 
			and numerics for $\delta >2$, including near the phase transition at $\delta_c^{\rm FM}=5.5$ (red dashed line).  
			The two diverge somewhat at smaller $\delta$ (later times), likely due to decoherence and 
			unmodeled systematic measurement errors.
			\textbf{c, d}, 
			Corresponding results for the AFM. For $P_z$, the $t$-MPS simulation accurately reproduces the experimental 
			data across the whole $\delta(t)$ sweep. For $\delta$ far above $\delta_c^{\rm AFM} = 0.6$ (blue dashed line), 
			there are many-body Rabi oscillations characteristic of the paramagnetic phase.
			\textbf{c}, Inset: ramp $\delta(t)$ used for the AFM simulation.
			\textbf{d}, Inset: zoom-in of lower left corner. At small $\delta$ (late times), the magnetization $m_{\rm AFM}^2$ 
			measured in experiment is below that predicted from the simulations. 
		}
		\label{fig:mps_exp_comparison}
	\end{figure*}
	%%%%%%%%%%%%%%%%%%%%%%%%%%%%%%%%%%%%%%
	
	The results of the $t$-MPS simulations are shown in Fig.~\ref{fig:mps_exp_comparison}, 
	which also includes direct comparisons to the experimental measurements, and to the DMRG ground state.
	For our ensemble of $N_{\rm dis} = 20$ independent $t$-MPS simulations, we show the average 
	values of these simulations with solid lines, while the shaded region indicates the standard deviation.
	
	Our first observable (Fig.~\ref{fig:mps_exp_comparison}a,c) is the staggered polarization 
	$P_z = \sum (\pm)_{A,B} \langle \sigma_i^z\rangle$.
	For the antiferromagnet, the agreement between the $t$-MPS simulations and 
	experiment is essentially perfect for all values of $\delta$.
	This is a strong indication that most dominant sources of error in the experiment have been accurately accounted for.
	For the ferromagnet, there is a small offset between the $t$-MPS calculation and the experimental 
	result at late times (small $\delta$).
	In particular, $P_z\to 0$ as $\delta\to 0$ for the $t$-MPS calculation, while $P_z \to -0.06$ in the experiment.
	This discrepancy is due to the sublattice-dependent depumping from the light-shift discussed in Sec.~\ref{SubSM:Decoherence}, 
	which we do not account for in the $t$-MPS simulations.
	
	As the state loses its initial $\sigma^z$ polarization, it concomitantly develops XY order. 
	This is tracked by the order parameter $m_{\rm FM}^2$ ($m_{\rm AFM}^2$ for the antiferromagnet), 
	shown in Fig.~\ref{fig:mps_exp_comparison}b,d.
	We obtain again a good agreement between the $t$-MPS simulation and the experiment at early times (large $\delta$), 
	although we caution that the initial positive value of $m_{\rm FM/AFM}^2=1/42$ is inherent to any $\sigma_i^z$-product state.
	On top of the smooth adiabatic envelope, the $t$-MPS simulations reveal coherent oscillations in 
	$P^z$ and $m_{\rm FM/AFM}^2$.
	These oscillations are a feature of the large-$\delta$ paramagnetic phase, 
	and are essentially Rabi oscillations between the classical N\'eel ground state and the 
	42-fold degenerate manifold of states with one spin-flip excitation.

	At small $\delta$, the experimental measurements of $m_{\rm FM/AFM}^2$ fall below the $t$-MPS predictions.
	This deficit likely arises from a combination of decoherence and unmodeled systematic errors, 
	such as experimental imperfections in the $\pi/2$-pulse rotation to the $x$ basis. 
	Regarding the latter, an imperfect basis rotation means that the operator measured in the
	experiment is not exactly $\sigma_i^x$ but some small modification of it, $\tilde{\sigma}_i^x = U \sigma_i^x U^\dagger$. 
	In XY-ordered states, $\langle \sigma^x \sigma^x\rangle = \langle \sigma^y \sigma^y \rangle$ 
	correlations are typically much larger than any other two-body operators, especially at long distances.
	Measuring any modified $\tilde{\sigma}_{i}^x$ will then generically reduce the value of the 
	inferred magnetization, $\tilde{m}_{FM}^2 = 2\sum_{i,j}\langle \tilde{\sigma}_i^x \tilde{\sigma}_j^x \rangle$.

	%%%%%%%%%%%% Figure SM 8 %%%%%%%%%%%%%%%%%%%
	\begin{figure}[t]
		\centering
		\includegraphics{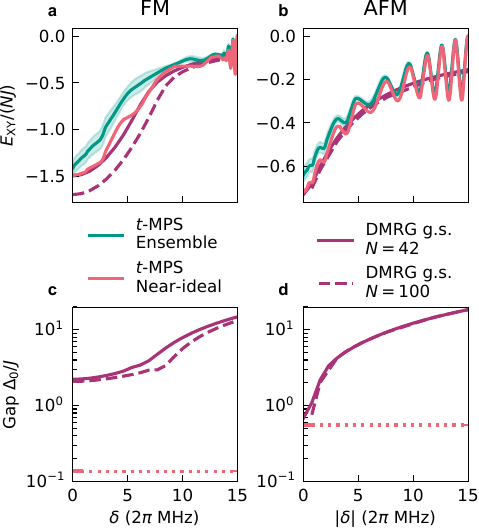}
		\caption{\textbf{Energetics of the simulated adiabatic preparation.}
			\textbf{a, b},  Interaction energy density $E_{\rm XY}/\bar{N}(\delta)$ in the $t$-MPS simulations of the $6 \times 7$ lattice. 
			The teal line and envelope are the disorder ensemble average and standard deviation, respectively. 
			Following a single state with minimal initialization errors (pink line), we see that $E_{\rm XY}$ tightly 
			follows the DMRG ground state value (purple), confirming that diabatic errors are negligible.
			\textbf{c, d}, Energy gaps $\Delta_0$ between the ground state and the first excited state in the $S^z=0$ sector, 
			obtained from DMRG. For the near-ideal initial state, the final energy density (pink dotted line) falls below the gap in both the FM and AFM case.}
		\label{fig:MPOMPS_energy}
	\end{figure}
	%%%%%%%%%%%%%%%%%%%%%%%%%%%%%%%%%%%%%%%

	We also use the $t$-MPS simulation to assess the quality of the adiabatic preparation.
	In particular, we are interested in how close the unitary dynamics comes to preparing the target ground state of $H_{\rm XY}$.
	We measure this via the XY energy $E_{\rm XY} = \langle H_{\rm XY}\rangle$, 
	which corresponds to the amount of energy the many-body state stores within the dipolar interaction.
	The ideal endpoint of the ramp is a state that maximizes $|E_{\rm XY}|$, 
	i.e. the ground state of $H_{\rm XY}$ 
	(or, the topmost state in the case of the negative-temperature preparation for the antiferromagnet). 
	We do not include measurement errors for this analysis, as we want to directly compare the 
	adiabatically prepared state to the ideal one.
	As a minor technical point, the ensemble of lattices used in the $t$-MPS 
	simulation occasionally have missing sites (representing an absence of Rydberg-excited atoms), 
	and always have some position disorder which modifies the couplings, $J a^3/r_{ij}^3$, 
	and hence the spectrum of $H_{\rm XY}$. 
	To treat the different lattices on even footing, when measuring $E_{\rm XY}$ 
	we  compute the expectation value of $H_{\rm XY}$ without position disorder in the couplings, 
	and normalize by the total number of active sites, $\bar{N}$, before taking the ensemble average. 
	
	Figures~\ref{fig:MPOMPS_energy}a,b show $E_{\rm XY}(\delta)/J\bar{N}$ 
	in the DMRG ground state (purple), the ensemble-averaged $t$-MPS simulation (teal), 
	and a single state within the $t$-MPS ensemble (pink) that had a nearly perfect initial 
	configuration: one missing site at the corner, and all remaining spins properly aligned with the staggered field. 
	Initially, the system is in a classical ensemble of $\sigma_i^z$-aligned product states, so $E_{\rm XY}(t=0) = 0$. 
	The dynamics generated by $H(t)$ produce the desired correlations among the spins; 
	the oscillations in $E_{\rm XY}$ at large $\delta$ are the paramagnetic Rabi oscillations 
	also observed in $P_z$ and $m_x^2$.
	At the end of the ramp, the ensemble averages 
	are $E_{\rm XY}^{\rm FM}/(\bar{N}J) = -1.41(8)$ and $E_{\rm XY}^{\rm AFM}/(\bar{N}J) = -0.64(3)$, 
	which respectively correspond to $94\pm5\%$ and $89\pm4\%$ of the $N=42$ ground state value. 
	Remarkably, the near-ideal initial state produces a near-ideal final state, achieving 
	$99.7\%$ (FM) and $98.2\%$ (AFM) of the ground state energy density.
	This indicates that any diabatic errors during the ramp are negligible compared 
	to the initialization errors. 
	
	As discussed in Sec.~\ref{SubSM:adiabatic_alternative}, the quality of a finite-time 
	adiabatic ramp crucially depends on the size of the many-body energy gap.
	For the $U(1)$-symmetric ramp at hand, the quantity is the (spin-)neutral gap, $\Delta_0 = E_1(S^z=0) - E_0(S^z=0)$.
	In the paramagnetic phase, $\Delta_0 \sim \delta$, 
	while in the XY-ordered phase one expects $\Delta_0^{FM} \sim 1/\sqrt{N}$ and 
	$\Delta_0^{AFM} \sim 1/N$~\cite{maleevDipoleForcesTwodimensional1976, Peter2012}.
	The numerical value of $\Delta_0$ on finite-size systems can be computed in DMRG 
	by solving for the lowest-energy state orthogonal to the previously obtained ground state, in the same $S^z=0$ sector.
	We plot $\Delta_0(\delta)$ in  Fig.~\ref{fig:MPOMPS_energy}c,d for both the $N=42$ and $N=100$ clusters.
	The behavior of $\Delta_0(\delta)$ differs somewhat from that seen in Sec.~\ref{SubSM:adiabatic_alternative}, 
	due to a difference in boundary conditions (open instead of periodic). 
	Across the phase diagram, $\Delta_0(\delta)$ is fairly large, which helps to explain 
	the success of the adiabatic preparation: the ramp decay time scale, 
	$\tau = {1.45\;\hbar/J}$, is slower than (FM) or approximately equal to (AFM) 
	the inverse gap, $\Delta_0^{-1} = {0.45/J}$ (FM), ${1.47/J}$ (AFM). 
	The smaller gap for the antiferromagnet is a manifestation of its frustration, 
	and makes adiabatically preparing its ground state more difficult compared to $H_{\rm XY}^{\rm FM}$. 
	Comparing $\Delta_0$ to the excess energy the end of the ramp, we find that the near-ideal 
	initial state ends up with a total effective energy, $\mathcal{E} =   N E_{\rm XY}/ \bar{N}$,  
	below the many-body gap.
	The difference is remarkably large for the ferromagnet ($\mathcal{E}/\Delta_0^{\rm FM} = 0.06$), 
	implying a near-flawless adiabatic sweep, while the margin for the antiferromagnet 
	is much narrower ($\mathcal{E}/\Delta_0^{\rm AFM} = 0.81$).

	\subsection{Thermal phase diagram}\label{SM:Therm}

	%%%%%%%%%%% Figure SM 9 %%%%%%%%%%%%%%
	\begin{figure*}%[t]
		\centering
		\includegraphics{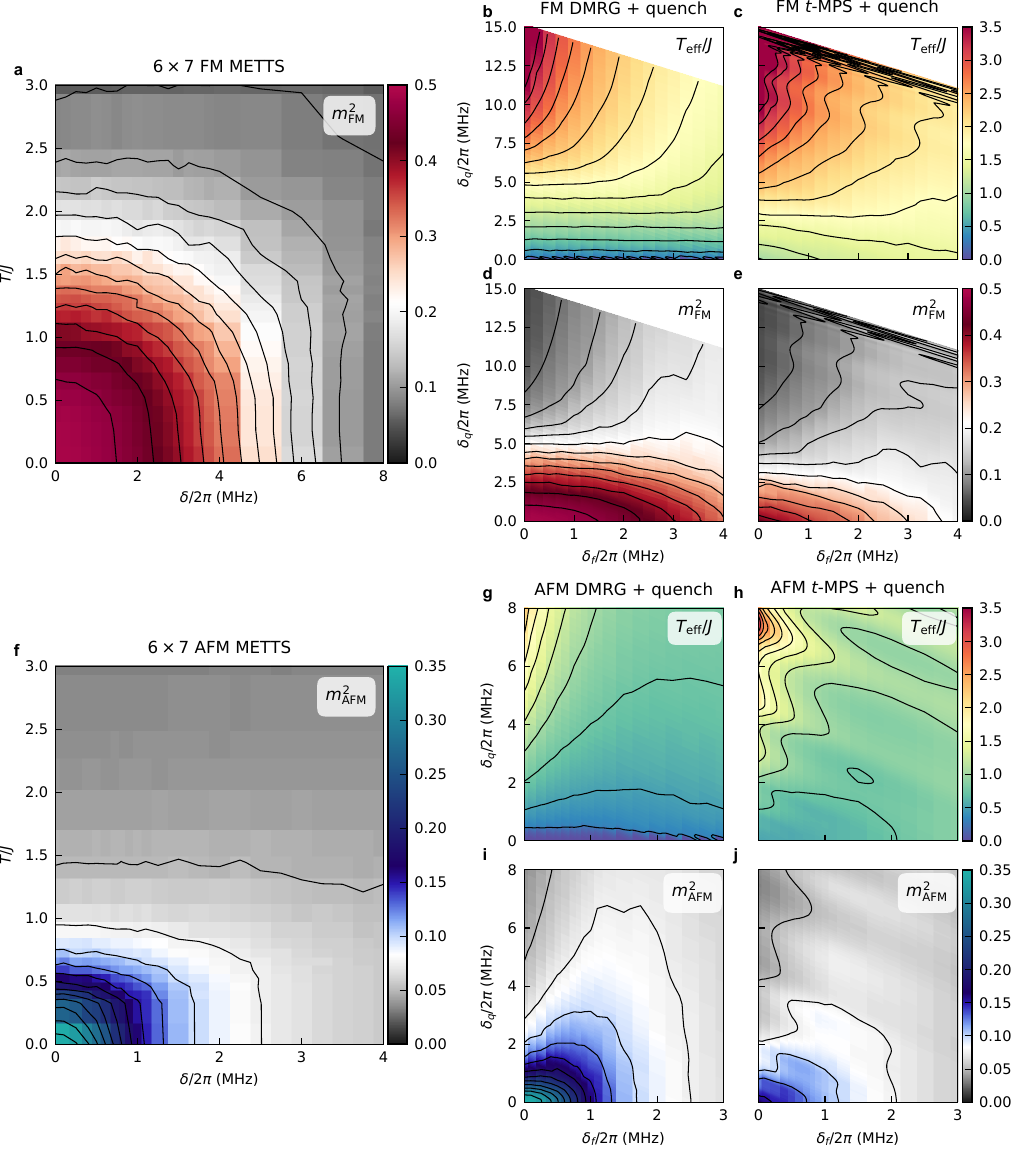}
		\caption{\textbf{Finite-temperature properties of $H_{\rm XY} +H_{\rm Z}$.} 
			\textbf{a},
			Phase diagram of $H_{\rm XY}^{\rm FM}+H_{\rm Z}$ at finite temperature $T$ and light-shift $\delta$, 
			computed from METTS on a $6\times 7$ array in the $M^z=0$ sector.
			We also include $T=0$ points calculated from DMRG. 
			The region with large magnetization $m_{\rm FM}^2$ 
			at small $\delta$ and small $T$ should correspond to the LRO phase in the thermodynamic limit. The colorbar is chosen so that dark red corresponds to the final $m_{\rm FM}^2$ calculated in the $t$-MPS simulation, absent measurement errors. Thin black lines are equal-magnitude contours to guide the eye.
			\textbf{b,c}, Estimated temperature of a quench experiment with final light-shift $\delta_f$ and quench magnitude $\delta_q$, taking the pre-quench configuration to be either the DMRG ground state (\textbf{b}) or the $t$-MPS ramp simulation ensemble (\textbf{c}).
			The oscillatory behavior seen in \textbf{c} stems from the paramagnetic Rabi oscillations discussed in Sec.\,\ref{SubSM:mps_results}.
			\textbf{d, e} Corresponding magnetization $m_{\rm FM}^2$ of the system at temperature $T_{\rm eff}(\delta_f,\delta_q)$.
			\textbf{f-j} Analogous results for the antiferromagnet.  The region with finite magnetization $m_{\rm AFM}^2$ is expected to become an algebraic-ordered (BKT) phase in the thermodynamic limit. 
		}
		\label{fig:temperature}
	\end{figure*}
	%%%%%%%%%%%%%%%%%%%%%%%%%%%%%%%%
	
	We conclude by discussing the phase diagram of $H_{\rm XY}+H_{\rm Z}$ at finite temperature, 
	$T$ (measured in unit of $k_B$).
	While two-dimensional, $U(1)$-symmetric systems can have XY LRO ground states, 
	for short-range interacting models such as $H_{\mathrm{nn}}$ this order 
	does not persist to finite temperature~\cite{merminAbsenceFerromagnetismAntiferromagnetism1966, 
		hohenbergExistenceLongRangeOrder1967a, merminCrystallineOrderTwo1968, frohlichAbsenceSpontaneousSymmetry1981a}.
	Physically, this is because spin-wave excitations (i.e. Goldstone modes) proliferate at finite temperature and destroy the XY order.  
	Instead, most two-dimensional XY models have an \textit{algebraic long-range ordered} 
	phase at low temperatures, separated from the high-$T$ disordered phase by a 
	Berezinskii-Kosterlitz-Thouless (BKT) transition at a critical temperature 
	$T_{\rm BKT}$~\cite{berezinskiiDestructionLongrangeOrder1971,
		berezinskiiDestructionLongrangeOrder1972, kosterlitzOrderingMetastabilityPhase1973, 
		kosterlitzCriticalPropertiesTwodimensional1974}.
	The low-$T$ phase is characterized by power-law-decaying correlations, $C^x(d)  \sim d^{-1/(2\pi K)}$, 
	with a temperature-dependent exponent $K$ that attains the universal value $K_{\rm BKT} = 2/\pi$ at $T_{\rm BKT}$. 
	For the classical nearest-neighbor XY model, $T_{\rm BKT}^{\rm cl}/(2 J) = 0.892943(2)$~\cite{tobochnikMonteCarloStudy1979a,
		uedaResolvingBerezinskiiKosterlitzThoulessTransition2021}, 
	while in the quantum spin-1/2 $H_{\rm nn}$ the transition is lowered to 
	$T_{\rm BKT}^{\rm nn}/(2J) =0.353(3)$~\cite{dingKosterlitzThoulessTransitionTwodimensional1990a,
		dingPhaseTransitionThermodynamics1992a}.
	
	Long-range ferromagnetic interactions can suppress the proliferation of spin-waves and thus renew the possibility for XY LRO 
	at finite temperature ~\cite{maleevDipoleForcesTwodimensional1976, frohlichPhaseTransitionsReflection1978, Peter2012}.
	With $1/r^\alpha$ ferromagnetic couplings, extensively large fluctuations of the spin orientation come at an energy cost proportional to $L^{4-\alpha}$, with $L$ the linear system size, so two-dimensional XY LRO can be thermodynamically stable when $\alpha \le 4$.
	Indeed, in 1976, Kunz and Pfister proved that the classical version of $H_{\rm XY}^{\rm FM}$
	exhibits a finite-temperature phase transition between the high-$T$ 
	disordered phase and a low-$T$ XY LRO phase~\cite{kunzFirstOrderPhase1976a}.
	Subsequent Monte Carlo simulations located this transition at 
	$T_{c}^{\rm FM}/(2J) = 3.96(4)$, and suggested it was weakly first-order~\cite{romanoComputerSimulationStudy1987a, romanoComputersimulationStudyDisordered1990}.
	We note this is contrary to the general expectation of a second-order symmetry-breaking transition, in a mean-field-like universality class when $\alpha \le 3$~\cite{Defenu2021,fisherCriticalExponentsLongRange1972,sakRecursionRelationsFixed1973}.
	Finally, $1/r^\alpha$ antiferromagnetic interactions do not essentially modify the energy of long-wavelength fluctuations, so one expects the low-$T$ physics of $H_{\rm XY}^{\rm AFM}$ to be similar to that of $H_{\rm nn}$~\cite{brunoAbsenceSpontaneousMagnetic2001a}.

	\subsubsection{Numerical phase diagram for \texorpdfstring{$N=42$}{N=42}}\label{SubSM:METTS}

    For a quantitative understanding of the thermal physics accessible in the experiment, we numerically investigate the finite temperature phase diagram for both the FM and the AFM on the $6\times 7$ lattice.
    While $H_{\rm XY}^{\rm FM}$ is amenable to Quantum Monte Carlo techniques, these are not an option for $H_{\rm XY}^{\rm AFM}$, which exhibits a sign problem. 
    Instead, for both we employ the Minimally Entangled Typical Thermal States (METTS) algorithm~\cite{stoudenmireMinimallyEntangledTypical2010}.
    This is a Markov chain Monte Carlo (MCMC) approach that alternates between evolving a state in imaginary time to inverse temperature $\beta/2$, and then taking a projective measurement as the initialization for the next imaginary time evolution. 
    The result is an ensemble of pure states, $\{ \ket{\psi_{\rm METTS}} \}$, that approximates the thermal density matrix $\rho \propto e^{-\beta H}$:   for any operator $\mathcal{O}$, the ensemble average of $\bra{\psi_{\rm METTS}} \mathcal{O} \ket{\psi_{\rm METTS}}$ approaches the thermal equilibrium value $\mathrm{Tr}[\rho\mathcal{O}]$.

    Due to the $U(1)$ symmetry, the thermal density matrix factorizes into a direct sum over the different magnetization sectors, $\rho= \bigoplus_{m=-N}^{N}\rho_{m}$. 
	Here, we sample only from the $m=0$ sector, as this is the most relevant one for the partial quench experiment. 
	For numerical convenience, we also truncate the long-range interactions to $R_{\rm max} < 3.7$, and omit the van der Waals coupling, position disorder, and the possibility of holes.
	We perform the imaginary time evolution using the same $W_{II}$ MPO-MPS method as in Sec.~\ref{SubSM:mps_results}, taking an MPS bond dimension of $\chi=256$. 
	We found very similar results using $\chi=128$ (not shown), albeit with some small quantitative shifts near the finite-temperature phase transition.
	To reduce sample autocorrelations, each projective measurement is made in a random basis determined by a depth-two, $U(1)$-conserving random unitary circuit~\cite{binderSymmetricMinimallyEntangled2017}. 
	By a standard blocking analysis, we estimate the resulting autocorrelation time to be about 10 MCMC steps~\cite{gubernatisQuantumMonteCarlo2016}.
	We therefore allow a warm-up time of 20 steps, and then generate 100-300 samples for each value of $\delta$ and $\beta$.

	In Fig.\,\ref{fig:temperature}a,f, we show 2D color plots of the squared magnetization $m_{\rm FM/AFM}^2$ at finite $T$ and  $\delta$.
	For the ferromagnet (Fig.\,\ref{fig:temperature}a), we observe a lobe around 
	$(T,\delta)=(0,0)$ that corresponds to the XY-ordered phase.
	The order begins to disappear around $T / J = 1.5$ (for the thermal phase transition) 
	and $\delta/J = 5$  (for the quantum phase transition). 
	Examining  $m_{\rm AFM}^2$ for the AFM case 
	(Fig.\,\ref{fig:temperature}b), we observe a smaller lobe with apparent XY order. 
	Although $H_{\rm XY}^{\rm AFM}$ is not predicted to host true long range order at $T>0$, 
	obtaining $m_{\rm AFM}^2 > 0$ is still possible on finite-size systems.

    Owing to the small system size, there is a smooth crossover between the ordered and disordered regimes for both models,
	and it is difficult to ascertain what the nature of the phase transition may be in the thermodynamic limit. 
	It should be possible to study larger system sizes for $H_{\rm XY}^{\rm FM}$ 
	using Quantum Monte Carlo methods~\cite{syljuasenQuantumMonteCarlo2002},
	which is beyond the scope of this work.
	For now, we cautiously estimate  $T_{\rm{XY}}^{\rm FM}/J \approx 1.5$ and 
	$ T_{\rm{XY}}^{\rm AFM}/J \approx 0.5$, as the $\delta=0$ crossover temperature into the high-$T$ phase. 
	Compared to $H_{\rm nn}$,  for which $T_{\rm BKT}^{\rm nn}/J = 0.706(6)$
	\cite{dingKosterlitzThoulessTransitionTwodimensional1990a, dingPhaseTransitionThermodynamics1992a}, 
	the dipolar ferromagnet appears to have a higher transition temperature 
	(although not as high as the classical model~\cite{romanoComputerSimulationStudy1987a,
		romanoComputersimulationStudyDisordered1990}), while the antiferromagnet may have a slightly lower one.
	
	\subsubsection{Temperature estimate of the final state}\label{SubSM:TempCal1}

	With our METTS representation of the thermal density matrix, we also determine the 
	temperature and $\delta$ dependence of the internal energy, $E(T) = \rm{Tr}[\rho H]$.
	The inverse function $T(E)$ defines a temperature calibration: we estimate the 
	effective temperature of a state from its energy density. 
	Inputting the mean final energy density of our $t$-MPS simulations (Sec.~\ref{SM:Adiabatic}), 
	we estimate the effective temperatures at the end of the adiabatic ramp to be 
	$T_{\rm MPS}^{\rm FM}/J = 0.95$ and $T_{\rm MPS}^{\rm AFM}/J = 0.53$.
	The $t$-MPS disorder ensemble results in a spread of energies $E_{\rm XY} \pm \sigma_E$; 
	the corresponding temperature intervals are $T_{\rm MPS}^{\rm FM} \in [0.46, 1.17]$ 
	and $T_{\rm MPS}^{\rm AFM}\in [0.45, 0.60]$.
	These intervals are asymmetric about the mean value due to the nonlinearity of $T(E)$.
	The obtained $T_{\rm MPS}^{\rm FM}$ appears to be below the estimated crossover temperature $T_{\rm XY}^{\rm FM}$ , while for the antiferromagnet $T_{\rm MPS}^{\rm AFM}$ is very close to the phase transition. 
	This is consistent with the wide spread in magnetizations $m_{\rm AFM}^2$ over the $t$-MPS ensemble, shown in Fig.\,\ref{fig:mps_exp_comparison}d.

	\subsubsection{Temperature calibration of quantum quenches}\label{SubSM:TempCal2}

	Performing an analogous $T(E)$ calibration at finite $\delta$, we also 
	estimate the effective temperatures produced by the quantum quench
	experiments (main text Fig.~\ref{fig:fig1}c,e,f),
	with final light-shift $\delta_f$ and quench magnitude $\delta_q$.
	We assume that, following the quench, the system equilibrates to a thermal state; extensively testing this assumption with numerical quench simulations is challenging, but may be interesting to explore in the future. 
	Barring the possibility of a nonthermal equilibrium, our basic expectation is that the quench affects the XY order by a mechanism not unlike a finite-temperature bath. 
	In particular, the excess energy added into the system should excite the low-energy, symmetry-restoring spin waves~\cite{calabreseTimeDependenceCorrelation2006}.
	If the resulting population density of spin waves at equilibrium is not too different from a true thermal distribution, then in the thermodynamic limit it will destabilize the XY AFM order but not the XY FM order at low temperature.   
	
	We first calculate the effective temperature assuming \textit{perfect} adiabatic preparation up to 
	the pre-quench point $\delta_f + \delta_q$, i.e. by evaluating the energy 
	$\langle H_{\rm XY} + H_{\rm Z} \left( \delta_f \right)\rangle$ in the DMRG ground state 
	of $H_{\rm XY} + H_{\rm Z} \left( \delta_f + \delta_q \right)$ and then converting it to a temperature.
	Figure~\ref{fig:temperature}b,g shows the effective temperature $T_{\rm eff}(\delta_f, \delta_q)/J$ for the FM and the AFM. 
	In the FM, modest quenches $\delta_q/2\pi < 4$ MHz uniformly increase the effective temperature as a function of $\delta_f$ over the range $\delta_f/2\pi \in [0, 3.5]$ MHz probed in the experiment.
	With larger quenches, the effective temperature increases rapidly for small values of $\delta_f$ and slows down at larger values of $\delta_f$.
	In the AFM, the effective temperature produced by even small quenches $\delta_q$ has a strong dependence on $\delta_f$, again being much more effective at raising the temperature as $\delta_f\to 0$ (i.e. the isotherms are steeply sloped at small $\delta_f$). 
	Figure~\ref{fig:temperature}d,i show the corresponding magnetization $m_{\rm FM/AFM}^2$ expected at $T_{\rm eff}(\delta_f, \delta_q)$.  
	Notably, in the AFM the large variation in $T_{\rm eff}(\delta_f)$ at a fixed $\delta_q$ leads to a ``tilted Matterhorn'' shape for the ordered region. 
	
	Finally, we estimate the effective temperature of the full experimental protocol by using the states produced in the $t$-MPS ramp simulation as the pre-quench configuration.  
	We show $T_{\rm eff}(\delta_f,\delta_q)/J$ for the FM and AFM in Fig.\,\ref{fig:temperature}c,h.
	As a consequence of the paramagnetic Rabi oscillations discussed previously in Sec.,\ref{SubSM:mps_results}, $T_{\rm eff}$ is also oscillatory. 
	For the FM, these oscillations only manifest at large $\delta_q$ (corresponding to pre-quench states taken very early in the ramp), while for the AFM they are relevant across the phase diagram. 
	The latter behavior ultimately stems from the fact that $\delta_c^{\rm AFM}/2\pi \approx 0.7$ MHz, so most pre-quench states are in the paramagnetic phase.
	
	The corresponding magnetization $m^2_{\rm FM/AFM}(T_{\rm eff})$ is shown in Fig.\,\ref{fig:temperature}e,j. 
    Comparing to the experimental results in Fig.~\ref{fig:fig1}e,f, we see that some qualitative features are reproduced by this calculation, especially for the AFM.
    For instance, the sloped phased boundary seen in the experiment at small $\delta_f$ is  due to the diagonal isotherms. 
    The calculation seems to differ from the experiment in the region with large $\delta_f$ (i.e. $\delta_f/2\pi > 2(1)$ MHz for the FM (AFM) ) and small $\delta_q$. 
	In particular, the order-disorder crossover appears to happen at larger $\delta_f$ than seen in the experiment, and the observed non-monotonic behavior of $m^2_{\rm FM/AFM}(\delta_q)$ is also less apparent. 
	These differences may come from the same unmodeled imperfections that led to a discrepancy in the absence of any quench (see Sec.,\ref{SubSM:mps_results}).
	Another possibility is that the thermal density matrix $\rho(T_{\rm eff})$ in the $M^z=0$ sector may be an inadequate approximation of the post-quench state, either due to nonthermal equilibration or neglected contributions from different magnetization sectors.

\end{document}